\documentclass[12pt]{article}
\newcommand{\beq}{\begin{equation}}
\newcommand{\eeq}{\end{equation}}
\newcommand{\bea}{\begin{eqnarray}}
\newcommand{\eea}{\end{eqnarray}}

\usepackage{epsf}
\usepackage{latexsym,amssymb,euscript}
\usepackage[dvips]{graphicx}
\usepackage{amsmath}

\makeatletter
\def\appendix{\par\clearpage
  \setcounter{section}{0}
  \setcounter{subsection}{0}
  \@addtoreset{equation}{section}
  \def\@sectname{Appendix~}
  \def\theequation{\thesection.\arabic{equation}}
  \def\theequation{\thesection.\arabic{equation}}
  \def\thesection{\Alph{section}}}
\makeatother

\begin{document}
\begin{titlepage}

\begin{center}
{\LARGE \bf
Inclusive production of a pair of hadrons separated
by a large interval of rapidity in proton collisions}
\end{center}

\vskip 0.5cm

\centerline{D.Yu.~Ivanov$^{1\P}$ and A.~Papa$^{2\dagger}$}

\vskip .6cm

\centerline{${}^1$ {\sl Sobolev Institute of Mathematics and Novosibirsk State
University,}}
\centerline{\sl 630090 Novosibirsk, Russia}

\vskip .2cm

\centerline{${}^2$ {\sl Dipartimento di Fisica, Universit\`a della Calabria,}}
\centerline{\sl and Istituto Nazionale di Fisica Nucleare, Gruppo collegato di
Cosenza,}
\centerline{\sl I-87036 Arcavacata di Rende, Cosenza, Italy}

\vskip 2cm

\begin{abstract}
We consider within QCD collinear factorization the inclusive process $p+p\to 
h_1+h_2+X$,
where the pair of identified hadrons, $h_1,h_2$, having large transverse
momenta is produced in high-energy proton-proton collisions. In particular,
we concentrate on the kinematics where the two identified hadrons in the final
state are separated by a large interval of rapidity $\Delta y$. In this case
the (calculable) hard part of the reaction receives large higher order
corrections $\sim \alpha^n_s \Delta y^n$. We provide a theoretical input
for the resummation of such contributions with next-to-leading logarithmic
accuracy (NLA) in the BFKL approach. Specifically, we calculate in NLA the
vertex (impact-factor) for the inclusive production of the identified hadron.
This process has much in common with the widely discussed Mueller-Navelet
jets production and can be also used to access the BFKL dynamics at proton
colliders. Another application of the obtained  identified-hadron vertex
could be the NLA BFKL description of inclusive forward hadron production
in DIS. 
\end{abstract}


$
\begin{array}{ll}
^{\P}\mbox{{\it e-mail address:}} &
\mbox{d-ivanov@math.nsc.ru}\\
^{\dagger}\mbox{{\it e-mail address:}} &
\mbox{papa@cs.infn.it}\\
\end{array}
$

\end{titlepage}

\vfill \eject

\section{Introduction}

The BFKL approach~\cite{BFKL} is the most suitable framework for the
theoretical description of the high-energy limit of hard or semi-hard
processes, i.e. processes where a hard scale exists that justifies
the application of perturbative QCD.

At high squared center of mass energy $s$ and for squared momentum
transfered $t$ not growing with $s$, large logarithms of the energy
compensate the small coupling and must be resummed at all orders of the
perturbative series. The BFKL approach provides a systematic way to
perform the resummation in the leading logarithmic approximation (LLA), which
means resummation of all terms $(\alpha_s\ln(s))^n$, and in the
next-to-leading logarithmic approximation (NLA), which means resummation of
all terms $\alpha_s(\alpha_s\ln(s))^n$.

In the BFKL approach, both in the LLA and in the NLA, the high-energy
scattering amplitudes are expressed by a suitable factorization of a
process-independent part, the Green's function for the interaction of two
Reggeized gluons, and process-dependent terms, the so-called impact factors
of the colliding particles~(see, for instance,~\cite{FF98}).

The Green's function is determined through the BFKL equation, which is
an iterative integral equation, whose kernel is known at the next-to-leading
order (NLO) both for forward scattering (i.e. for $t=0$ and color singlet
in the $t$-channel)~\cite{FL98,CC98} and for any fixed (not growing with
energy) momentum transfer $t$ and any possible two-gluon color state in the
$t$-channel~\cite{FF05}.

The impact factors of the colliding particle are a necessary ingredient
for the complete description of a process in the BFKL approach and, therefore,
to get a contact with phenomenology. The only impact factors calculated
so far with NLO accuracy are those for colliding quark and
gluons~\cite{FFKP99,FFKP99g,Cia}, for forward jet
production~\cite{Bartels:2002yj,Caporale:2011cc,Ivanov:2012ms},
for the $\gamma^* \to \gamma^*$ transition~\cite{gammaIF} and for the
$\gamma^*$ to light vector meson transition ~\cite{IKP04}.
 Recently, the NLO $\gamma^* \to \gamma^*$ impact factor has been calculated
in the coordinate representation \cite{Balitsky:2010ze}. 

The impact factors for colliding partons~\cite{FFKP99,FFKP99g,Cia} played a
role in the proof of fulfillment of bootstrap conditions for the gluon
Reggeization in the NLA~\cite{FF98} and are at the basis of the
calculation of the impact factors for forward jets.

The impact factor for the $\gamma^* \to \gamma^*$ transition is certainly
the most important one from the phenomenological point of view, since it
enters the prediction of the $\gamma^* \gamma^*$ total cross section,
which is believed to be the golden channel for the manifestation of the BFKL
dynamics~\cite{gold}. Its calculation turned out to be very complicated
and took several year to be completed~\cite{gammaIF}.
A relatively simpler calculation led to the leading-twist NLO impact factor
for the transition from a virtual photon $\gamma^*$ to a light neutral vector
meson $V=\rho^0, \omega, \phi$, which was used to study the
$\gamma^* \gamma^* \to V V$ reaction~\cite{mesons}. However, both for the
$\gamma^* \gamma^*$ total cross section and for the production of two
light vector mesons in $\gamma^* \gamma^*$ collisions, accurate experimental
data will possibly come only from future high-energy $e^+e^-$ and $e\gamma$
colliders.

On the other side, the availability of the NLO forward jet impact
factors~\cite{Bartels:2002yj} allowed the NLA
calculation~\cite{Colferai:2010wu} of the cross-section for the production of
Mueller-Navelet jets~\cite{Mueller:1986ey}, which can be tested at
hadron colliders, such as Tevatron and LHC.

In the present work we calculate in the NLO a new impact factor, namely the
one for the forward production of an identified hadron, which can be used to
determine the cross section for the semi-inclusive process $p+p\to h_1+h_2+X$,
in the kinematics where the identified hadrons in the final state, $h_1$,
$h_2$, have large transverse momenta and are separated by a large interval of
rapidity $\Delta y$. This process has much in common with the widely discussed
Mueller-Navelet jets production and can be also used to access the BFKL
dynamics at proton colliders.
Very similarly to the case of forward jet production,
the calculation of this impact factor lies at the border between collinear
and BFKL factorization: the proton in the initial state emits a parton
which, in its turn, produces the hadron in the final state. This calls for
the collinear factorization of the hard part of the process with parton
distribution functions (PDFs) in the initial-state proton and with the relevant
fragmentation functions (FFs) for the production of the final-state hadrons,
both functions obeying the standard DGLAP evolution~\cite{DGLAP}.

The paper is organized as follows. In the next Section we will present the
factorization structure of the cross section, recall the definition
of BFKL impact factor and discuss the treatment of the divergences arising
in the calculation; in Section~3 we give the derivation of LO impact factor;
Section~4 is devoted to the calculation of the NLO impact factor; finally,
in Section~5 we summarize our results.

\section{General framework}

We consider the process
\beq
p(p_1)+p(p_2)\to h_1(k_1)+h_2(k_2)+X \
\label{process}
\eeq
in the kinematical region where both identified hadrons have large transverse
 momenta\footnote{See Eq.~(\ref{sudakov}) below for the definition
of the transverse part of a 4-vector.},
$\vec k_{1}^2\sim \vec k_{2}^2 \gg \Lambda_{\rm QCD}^2$.
This provides the hard scale, $Q^2\sim \vec k_{1,2}^2$, which makes
perturbative QCD methods applicable. Moreover, the energy of the proton
collision is assumed to be much bigger than the hard scale,
$s=2p_1\cdot p_2\gg \vec k_{1,2}^2$.

We consider the leading behavior in the $1/Q$-expansion (leading twist
approximation). With this accuracy one can neglect the masses of initial
protons and identified hadrons. The state of the identified hadrons can be
described completely by their (pseudo)rapidities $y_{1,2}$\footnote{For
massless particle the rapidity coincides with pseudorapidity, $y=\eta$,
the latter being related to the particle polar scattering angle by
$\eta=-\ln\tan \frac{\theta}{2}$.} and transverse momenta $\vec k_{1,2}$.

In QCD collinear factorization the cross section of the process reads
\beq
\frac{d\sigma}{dy_1dy_2d^2\vec k_1d^2\vec k_2} =\sum_{i,j=q,g}\int\limits^1_0
\int\limits^1_0 dx_1dx_2 f_i(x_1,\mu) f_j(x_2,\mu)
\frac{d\hat \sigma(x_1 x_2 s,\mu)}{dy_1dy_2d^2\vec k_1d^2\vec k_2}\;,
\label{ff}
\eeq
where the $i,j$ indices specify parton types, $i,j=q,\bar q, g$, $f_i(x,\mu)$
denotes the initial protons parton density functions (PDFs), the longitudinal
fractions of the partons involved in the hard subprocess are $x_{1,2}$,
$\mu$ is the factorization scale and $d\hat \sigma(x_1 x_2 s,\mu)$ is the
partonic cross section for the production of identified hadrons. The
latter is expressed in terms of parton fragmentation functions (FFs), to be
specified later.

It is convenient to define the Sudakov decomposition for the identified-hadron
momentum,
\beq
k_h= \alpha_h p_1+ \frac{\vec k_h^2}{\alpha_h s}p_2+k_{h\perp} \ , \quad
k_{h\perp}^2=-\vec k_h^2 \ ,
\label{sudakov}
\eeq
where the longitudinal fraction $\alpha_h$ is related to the hadron
rapidity as $y=\frac{1}{2}\ln\frac{\alpha_h^2 s}{\vec k_h^2}$,
$dy=\frac{d\alpha_h}{\alpha_h}$ in the center of mass system.

Another important longitudinal fraction is related with the collinear
fragmentation of the parton $i$ into a hadron $h$. The probability
for such fragmentation is generically expressed as the convolution of a
parton fragmentation function $D^h_i$ and a coefficient function $C^h_i$,
\beq
d\sigma_i =C^h_i(z) dz\to d\sigma^h =d\alpha_h\int\limits^1_{\alpha_h}
\frac{dz}{z} D^h_i\left(\frac{\alpha_h}{z},\mu\right) C^h_i(z,\mu) \ .
\label{fragm-r}
\eeq
Here $C^h_i$ is the cross section (calculable in QCD perturbation theory) for
the production of a parton with momentum fraction $z$; the non-perturbative,
large-distance part of the transition to a hadron with momentum fraction
$\alpha_h$, is described in terms of the fragmentation function $D^h_i$;
$\mu$ stands here for the factorization scale.

Thus, the cross section $d\hat \sigma(x_1 x_2 s,\mu)$ in Eq.~(\ref{ff}) can be
further presented as the convolution of two fragmentation functions with the
cross section of the partonic subprocess
\beq
i(x_1 p_1)+j(x_2 p_2)\to m(k_1/z_1)+ n(k_2/z_2)+X \ ,
\eeq
where $X$ stands for inclusive production of additional, ``unidentified''
partons. This partonic cross section is free from infrared divergences, but it
contains the collinear singularities which are absorbed into the definition of
PDFs and FFs, thus leading to a finite result for our process of
interest~(\ref{process}) in collinear factorization.

In the case of large interval of rapidity between the two identified hadrons,
the energy of the partonic subprocess is much larger than hadron transverse
momenta, $x_1x_2 s\gg \vec k^2$. In this region the perturbative
partonic cross section receives at higher orders large contributions
$\sim \alpha^n_s\ln^n \frac{s}{\vec k^2}$, related with large energy
logarithms. It is the aim of this paper to elaborate the resummation of such
enhanced contributions with NLA accuracy using the BFKL approach.

Let us remind some generalities of the BFKL method. Due to the optical theorem,
the cross section is related to the imaginary part of the forward
proton-proton scattering amplitude,
\beq
\sigma =\frac{{\cal I}m_s A}{s} \ .
\eeq
In the BFKL approach the kinematic limit $s\gg \vec k^2$ of the
forward amplitude may be presented in $D$ dimensions as follows:
\beq
{\cal I}m_s
\left( {\cal A} \right)=\frac{s}{(2\pi)^{D-2}}\int\frac{d^{D-2}\vec q_1}{\vec
q_1^{\,\, 2}}\Phi_1(\vec q_1,s_0)\int
\frac{d^{D-2}\vec q_2}{\vec q_2^{\,\,2}} \Phi_2(-\vec q_2,s_0)
\int\limits^{\delta +i\infty}_{\delta
-i\infty}\frac{d\omega}{2\pi i}\left(\frac{s}{s_0}\right)^\omega
G_\omega (\vec q_1, \vec q_2)\, ,
\eeq
where the Green's function obeys the BFKL equation
\beq
\omega \, G_\omega (\vec q_1,\vec q_2)  =\delta^{D-2} (\vec q_1-\vec q_2)
+\int d^{D-2}\vec q \, K(\vec q_1,\vec q) \,G_\omega (\vec q, \vec q_1) \;.
\eeq
The energy scale parameter $s_0$ is arbitrary, the amplitude, indeed,
does not depend on its choice within NLA accuracy due to the properties of
NLO impact factors $\Phi_{1,2}$ to be discussed below.

What remains to be calculated are the NLO impact factors $\Phi_1$ and
$\Phi_2$ which describe the inclusive production of the identified hadrons
$h_1$ and $h_2$, with fixed transverse momenta $\vec k_1$, $\vec k_2$ and
rapidities $y_1$, $y_2$, in the fragmentation regions of the colliding protons
with momenta $p_1$ and $p_2$, respectively. For definiteness, we will consider
the case when the identified hadron belongs to the fragmentation region of
the proton with momentum $p_1$, i.e. the hadron is produced in the collision
of the proton with momentum $p_1$ off a Reggeon with incoming (transverse)
momentum $q$.

Technically, this is done using as starting point the definition of
inclusive parton impact factor, given in Refs.~\cite{FFKP99,FFKP99g}, for the
cases of incoming quark(antiquark) and gluon, respectively. This definition
involves the integration over all possible intermediate
states appearing in the parton-Reggeon collision, see Fig.~1. Up to the
next-to-leading order, this means that we can have one or two partons in the
intermediate state.
Here we review the important steps and give the formulae for the LO parton
impact factors.

Note that both the kernel of the equation for the BFKL Green's function and
the parton impact factors can be expressed in terms of the gluon Regge
trajectory,
\begin{equation}
j(t)\;=\;1\:+\:\omega (t)\;,
\end{equation}
and the effective vertices for the Reggeon-parton interaction.

To be more specific, we will give below the formulae for the case of forward
quark impact factor considered in $D=4+2\epsilon$ dimensions of dimensional
regularization.
We start with the LO, where the quark impact factors are given by
\begin{equation}
\Phi _{q}^{(0)}(\vec q\,)\;=\;\sum_{\{a\}}\:\int \:\frac{dM_a^2}{2\pi }
\:\Gamma_{a q}^{(0)}(\vec q\,)\:[\Gamma_{a q}^{(0)}(\vec q\,)]^*
\:d\rho_a\;,
\label{eq:a7}
\end{equation}
where $\vec q$ is the Reggeon transverse momentum, and $\Gamma ^{(0)}_{a q}$
denotes the Reggeon-quark vertices in the LO or Born approximation.
The sum $\{a\}$ is over all intermediate states $a$ which contribute to the
$q\rightarrow q$ transition. The phase space element $d\rho_a$ of the state
$a$, consisting of particles with momenta $\ell_n$, is ($p_q$ is the initial 
quark momentum)
\begin{equation}
d\rho _a\;=\;(2\pi )^D\:\delta^{(D)} \left( p_q+q-\sum_{n\in a}\ell
_n\right) \:\prod_{n\in a}\:\frac{d^{D-1}\ell _n}{(2\pi )^{D-1}2E_n}\;,
\label{eq:a8}
\end{equation}
while the remaining integration in~(\ref{eq:a7}) is over the squared
invariant mass of the state $a$,
\[
M_a^2\;=\;(p_q+q)^2\; .
\]

In the LO the only intermediate state which contributes is a one-quark state,
$\{a\}=q$. The integration in Eq.~(\ref{eq:a7}) with the known Reggeon-quark
vertices $\Gamma _{q q}^{(0)}$ is trivial and the quark impact factor reads
\begin{equation}
\Phi _{q}^{(0)}(\vec q\,
)\;=g^2 \frac{\sqrt{N^2-1}}{2N} \;,
\label{eq:a77}
\end{equation}
where $g$ is the QCD coupling, $\alpha_s=g^2/(4\pi)$, $N=3$ is the number of
QCD colors.

In the NLO the expression~(\ref{eq:a7}) for the quark impact factor has to be
changed in two ways. First, one has to take into account the radiative
corrections to the vertices,
$$
\Gamma _{q q}^{(0)}\to \Gamma_{q q}=\Gamma_{q q}^{(0)}+\Gamma _{q q}^{(1)} \;.
$$
Second, in the sum over $\{a\}$ in~(\ref{eq:a7}), we have to include more
complicated states which appear in the next order of perturbative theory. For
the quark impact factor this is a state with an additional gluon, $a=q g$.
However, the integral over $M_a^2$ becomes divergent when an extra gluon
appears in the final state. The divergence arises because the gluon may be
emitted not only in the  fragmentation region of initial quark, but also in
the central rapidity region. The contribution of the central region must be
subtracted from the impact factor, since it is to be assigned in the BFKL
approach to the Green's function. Therefore the result for the forward quark
impact factor reads
\begin{eqnarray}
\label{eq:a19}
\Phi _{q}(\vec q\, ,s_0)
=\left(\frac{ s_0}{\vec q^{\: 2} }\right)^{\omega(-\vec q^{\: 2})}
\:\sum_{\{a\}}\:\int \:\frac{dM_a^2}{2\pi }\:\Gamma
_{aq}(\vec q\,)\:[\Gamma _{aq}(\vec q\,)]^*\:d\rho _a\:\theta(s_\Lambda -M_a^2)
&&
\nonumber \\
-\frac{1}{2}\int d^{D-2} k\frac{\vec q^{\,\, 2}}{\vec k^{\, 2}}
\Phi_{q}^{(0)}(\vec k)\mathcal{K}_r^{(0)}(\vec k,\vec q\,)
\ln\left( \frac{s_{\Lambda}^2}{(\vec k-\vec q\,)^2 s_0}\right) \;.
&&
\end{eqnarray}
The second term in the r.h.s. of Eq.~(\ref{eq:a19}) is the subtraction
of the gluon emission in the central rapidity region. Note that, after this
subtraction, the intermediate parameter $s_\Lambda $ in the r.h.s. of
Eq.~(\ref{eq:a19}) should be sent to infinity.
The dependence on $s_\Lambda$ vanishes because of the cancellation between the
first and second terms. $K_r^{(0)}$ is the part of LO BFKL kernel related to
real gluon production,
\begin{equation}
\label{eq:a20}
K_r^{(0)}(\vec k,\vec q\,)\;=\;\frac{2 g^2N}{(2\pi )^{D-1}}
\frac{1}{(\vec k-\vec q\,)^2} \; .
\end{equation}
The factor in Eq.~(\ref{eq:a19}) which involves the Regge trajectory arises
from the change of energy scale~($\vec q^{\: 2}\to s_0$) in the vertices
$\Gamma$. The trajectory function $\omega (t)$ can  be taken here in the
one-loop approximation ($t=-\vec q^{\,\,2}$),
\begin{equation}
\omega (t)\;=\;\frac{g^2t}{(2\pi )^{D-1}}\frac{N}2\int \frac{d^{D-2}k}
{\vec k^2(\vec q-\vec k)^2}\;=\;-\;g^2N\frac{\Gamma (1-\varepsilon )}
{(4\pi )^{D/2}} \frac{\Gamma^2(\varepsilon )}{\Gamma (2\varepsilon )}
(\vec q^{\,\,2})^\varepsilon \; . \label{eq:b20}
\end{equation}

In the Eqs.~(\ref{eq:a7}) and (\ref{eq:a19}) we suppress for shortness the
color indices (for the explicit form of the vertices see~\cite{FFKP99}).
The gluon impact factor $\Phi_g(\vec q\,)$ is defined similarly. In the gluon
case only the single-gluon intermediate state contributes in the LO, $a=g$,
which results in
\begin{equation}
\Phi _{g}^{(0)}(\vec q\,
)\;=\frac{C_A}{C_F}\Phi _{q}^{(0)}(\vec q\,) \; ,
\label{eq:a77a}
\end{equation}
here $C_A=N$ and $C_F=(N^2-1)/(2N)$.  In the NLO  additional two-gluon,
$a=g g$, and quark-antiquark, $a=q \bar q$, intermediate states have to be
taken into account in the calculation of the gluon impact factor.

The identified hadron production vertex we want to calculate is simply related
with inclusive parton impact factors.
In order to allow the inclusive production of a given hadron,
one of these integrations in the definition of parton impact factors is
``opened'' (see Fig.~2). This means,
in practice, that the integration over the momentum of one of the
intermediate-state partons is replaced by the convolution with a suitable
fragmentation function. We illustrate the procedure starting from the
LO impact factor, contributing to the BFKL amplitude in the LLA, then we move
on to the NLO impact factor, relevant for the BFKL resummation in the NLA.

\begin{figure}[tb]
\centering
\includegraphics{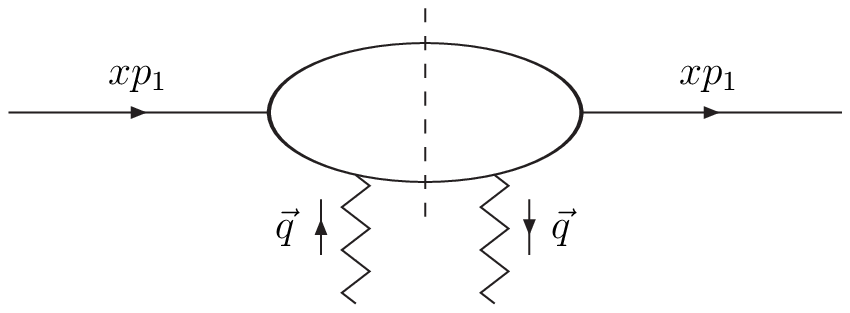}
\caption{Diagrammatic representation of the forward parton impact factor.}
\label{fig:if}
\end{figure}

\begin{figure}[tb]
\centering
\includegraphics[scale=0.8]{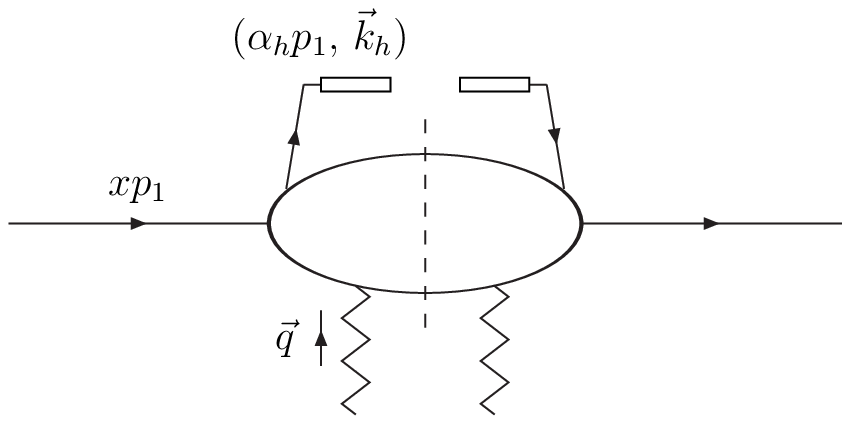}
\hspace{1cm}
\includegraphics[scale=0.8]{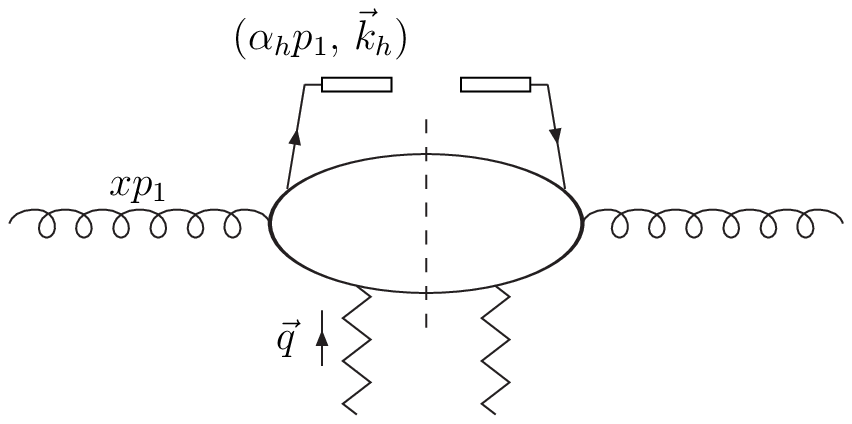}
\caption[]{Diagrammatic representation of the vertex for the identified
hadron production for the case of incoming quark (left) or gluon (right).}
\label{fig:vertex}
\end{figure}

\section{Impact Factor in the LO}

The inclusive LO impact factors of quark (antiquark) and gluon are given in
Eqs.~(\ref{eq:a77}) and~(\ref{eq:a77a}).
It is important that these expressions are valid also in
non-integer dimensions.

The inclusive LO impact factor of proton may be thought of as
the convolution of quark and gluon impact factors with the corresponding
proton PDFs,
\beq
d\Phi={\cal C} \, dx \left(\frac{C_A}{C_F}f_g(x)+\sum_{a=q,\bar q}
f_a(x)\right)\ , \quad {\cal C}=g^2\frac{\sqrt{N^2-1}}{2N}=2\pi\alpha_s
\sqrt{\frac{2\, C_F}{C_A}} \;.
\label{inclusive}
\eeq

\begin{figure}[tb]
\centering
\includegraphics[scale=0.8]{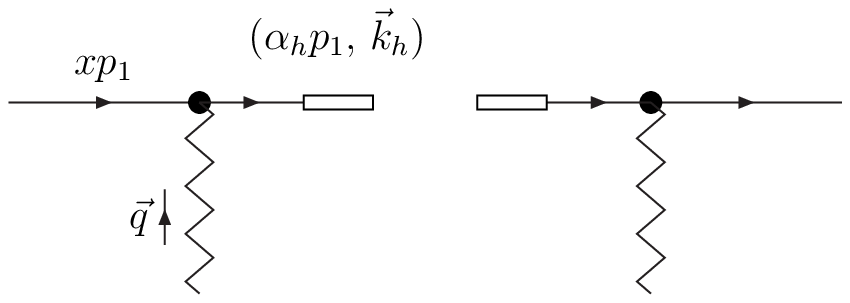}
\hspace{1cm}
\includegraphics[scale=0.8]{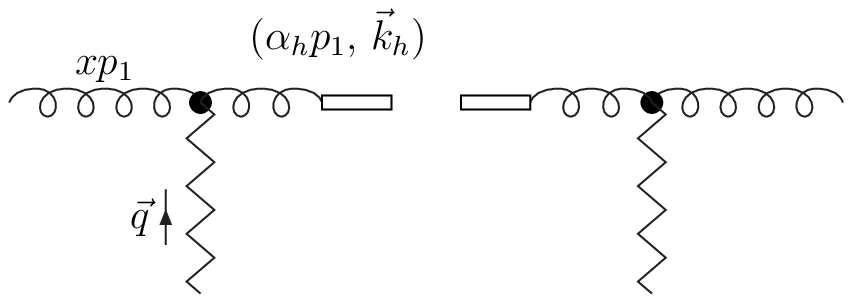}
\caption[]{Diagrammatic representation of the LO vertex for the identified
hadron production for the case of incoming quark (left) and gluon (right).}
\label{fig:vertex_LLA}
\end{figure}

In order to establish the proper normalization for the impact factor in the
case of fragmentation with an identified hadron, we insert into the inclusive
impact factor, Eq.~(\ref{inclusive}), a delta function and the integration over
the ``parent parton'' transverse momentum, $d^2\vec k$, and
use Eq.~(\ref{fragm-r}) to get:
\beq
\frac{d\Phi^h}{\vec q^{\,\, 2}}={\cal C} \,d\alpha_h \frac{d^2\vec k}
{\vec k^{\, 2}}\int\limits^1_{\alpha_h} \frac{dx}{x}
\, \delta^{(2)}\left(\vec k-\vec q\right)\left(\frac{C_A}{C_F}f_g(x)
D^h_g\left(\frac{\alpha_h}{x}\right)+\sum_{a=q,\bar q} f_a(x)
D^h_a\left(\frac{\alpha_h}{x}\right)\right) \ ,
\eeq
an expression for the LO impact factor schematically presented in
Fig.~\ref{fig:vertex_LLA}.

What is left is to express the ``parent parton'' variables by those of the
identified hadron, $\vec k=(x/\alpha_h) \vec k_h$ (note that
$d^2 \vec k_h/\vec k_h^2=d^2 \vec k/\vec k^2$).

Although the results for the fragmentation impact factors should be
expressed in terms of quantities describing the identified hadron,
$\alpha_h, \vec k_h$, it is convenient during the calculation to operate with
the kinematical variables of the parent parton, $\alpha, \vec k$
(in LO, $\alpha=x$).
The transition to the hadron variables may be easily done at the end of the
calculation.

In what follows we will calculate the projection of the impact factor
on the eigenfunctions of LO BFKL kernel, i.e. the impact factor
in the so called $(\nu,n)$-representation,
\beq
\Phi(\nu,n)=\int d^2\vec q \,\frac{\Phi(\vec q\,)}{\vec q^{\,\, 2}}\frac{1}{\pi
\sqrt{2}}\left(\vec q^{\,\, 2}\right)^{i\nu-\frac{1}{2}} e^{i n \phi} \;.
\label{nu_rep}
\eeq
Here $\phi$ is the azimuthal angle of the vector $\vec q$ counted from
some fixed direction in the transverse space.

\section{The NLO calculation}

We will work in $D=4+2\epsilon$ dimensions and calculate the NLO impact factor
directly in the $(\nu,n)$-representation~(\ref{nu_rep}), working out
separately virtual corrections and real emissions. To this purpose we introduce
the ``continuation'' of the LO BFKL eigenfunctions to non-integer dimensions,
\beq
\left(\vec q^{\,\, 2} \right)^{\gamma}e^{i n \phi}\to  \left(\vec q^{\,\,2}
\right)^{\gamma-{n \over 2}}\left(\vec q \cdot \vec l \,\, \right)^n\;,
\label{eigen}
\eeq
where $\gamma=i\nu-\frac{1}{2}$ and  $\vec l^{\:2}=0$. It is assumed that the
vector $\vec l$ lies only in the first two of the $2+2\epsilon$ transverse
space dimensions, i.e. $\vec l=\vec e_1+i\,  \vec e_2$, with 
$\vec e_{1,2}^{\:2}=
1$, $\vec e_{1}\cdot \vec e_2=0$. In the limit $\epsilon\to 0$ the r.h.s. of
Eq.~(\ref{eigen}) reduces to the LO BFKL eigenfunction.
This technique was used recently in Ref.~\cite{Kirschner:2009qu}. An even
more general method, based on an expansion in traceless products, was uses
earlier in Ref.~\cite{Kotikov:2000pm} for the calculation of NLO BFKL kernel
eigenvalues.

Thus, for the case of non-integer dimension the LO result for the impact
factor reads
\beq
\frac{d\Phi^h}{\vec q^{\,\, 2}}={\cal C} \,d\alpha_h \frac{d^{2+2\epsilon}
\vec k}{\vec k^{\, 2}}\int\limits^1_{\alpha_h} \frac{dx}{x}
\, \delta^{(2+2\epsilon)}\left(\vec k-\vec q\right)\left(\frac{C_A}{C_F}f_g(x)
D^h_g\left(\frac{\alpha_h}{x}\right)+\sum_{a=q,\bar q} f_a(x)
D^h_a\left(\frac{\alpha_h}{x}\right)\right)\;,
\eeq
which gives in the $(\nu,n)$-representation the result
\beq
\frac{\pi\sqrt{2}\, \vec k^{\, 2}}{{\cal C}}\frac{d\Phi^h(\nu,n)}
{d\alpha_h d^{2+2\epsilon}\vec k}=\int\limits^1_{\alpha_h}\frac{dx}{x}
\left( \frac{C_A}{C_F}f_g(x) D_g^h\left(\frac{\alpha_h}{x}\right)
+\sum_{a=q,\bar q}f_a(x) D_a^h\left(\frac{\alpha_h}{x}\right)\right)
\left(\vec k^{\,\,2} \right)^{\gamma-{n \over 2}}
\left(\vec k \cdot \vec l \,\, \right)^n \;.
\label{LO}
\eeq

Collinear singularities which appear in the NLO calculation are removed by the
renormalization of PDFs and FFs.
The relations between the bare and renormalized quantities are
\bea
&
f_q(x)=f_q(x,\mu_F)-\frac{\alpha_s}{2\pi}\left(\frac{1}{\hat \epsilon}
+\ln\frac{\mu_F^2}{\mu^2}\right)
\int\limits^1_x\frac{dz}{z}\left[P_{qq}(z)f_q(\frac{x}{z},\mu_F)
+P_{qg}(z)f_g(\frac{x}{z},\mu_F)\right] \;,
&\nonumber\\
&
f_g(x)=f_g(x,\mu_F)-\frac{\alpha_s}{2\pi}\left(\frac{1}{\hat \epsilon}
+\ln\frac{\mu_F^2}{\mu^2}\right)
\int\limits^1_x\frac{dz}{z}\left[P_{gq}(z)f_q(\frac{x}{z},\mu_F)
+P_{gg}(z)f_g(\frac{x}{z},\mu_F)\right]\;,
\label{DGLAPpdfs}
\eea
where $\frac{1}{\hat \epsilon}=\frac{1}{\epsilon}+\gamma_E-\ln (4\pi)\approx 
\frac{\Gamma(1-\epsilon)}{\epsilon (4\pi)^\epsilon}$, and the DGLAP kernels
are given by
\bea
P_{gq}(z)&=&C_F\frac{1+(1-z)^2}{z}\;,\\
P_{qg}(z)&=&T_R\left[z^2+(1-z)^2\right]\;,\\
P_{qq}(z)&=&C_F\left( \frac{1+z^2}{1-z} \right)_+
          = C_F\left[ \frac{1+z^2}{(1-z)_+} +{3\over 2}\delta(1-z)\right]\;,\\
P_{gg}(z)&=&2C_A\left[\frac{1}{(1-z)_+} +\frac{1}{z} -2+z(1-z)\right]
          +\left({11\over 6}C_A-\frac{n_f}{3}\right)\delta(1-z)\;,
\eea
with $T_R=1/2$. Here and below we always adopt the $\overline{\rm{MS}}$ scheme.
Similarly, for FFs we have
\bea
&D^h_q(x)=D^h_q(x,\mu_F)-\frac{\alpha_s}{2\pi}\left(\frac{1}{\hat \epsilon}
+\ln\frac{\mu_F^2}{\mu^2}\right)
\int\limits^1_x\frac{dz}{z}\left[D^h_q(\frac{x}{z},\mu_F)P_{qq}(z)
+D^h_g(\frac{x}{z},\mu_F)P_{gq}(z)\right]\;,&\nonumber\\
&D^h_g(x)=D^h_g(x,\mu_F)-\frac{\alpha_s}{2\pi}\left(\frac{1}{\hat \epsilon}
+\ln\frac{\mu_F^2}{\mu^2}\right)
\int\limits^1_x\frac{dz}{z}\left[D^h_q(\frac{x}{z},\mu_F)P_{qg}(z)
+D^h_g(\frac{x}{z},\mu_F)P_{gg}(z))\right]\;.&
\label{DGLAPffs}
\eea
Here $\mu$ is an arbitrary scale parameter introduced by the dimensional
regularization, which cancels out in the results for physical quantities.
Note the difference in the non-diagonal terms ($P_{qg}\leftrightarrow P_{gq}$)
between the formulas for PDFs' and FFs' renormalization, which is due to the
fact that in the parton to hadron splitting the hadron is in the first case
in the initial state, whereas in the second case it is in the final state.

Owing to $\vec k=(x/\alpha_h)\vec k_h$, one gets
\bea
&& \frac{\pi\sqrt{2}\, \vec k^{\, 2}}{{\cal C}}\frac{d\Phi^h(\nu,n)}
{d\alpha_h d^{2+2\epsilon}\vec k} =  \label{LO_bis} \\
&&\!\!\left(\vec k_h^{\,\,2}
\right)^{\gamma-{n \over 2}}\left(\vec k_h \cdot \vec l \,\, \right)^n
\!\!\int\limits^1_{\alpha_h}\frac{dx}{x}\left( \frac{C_A}{C_F}f_g(x)
D_g^h\left(\frac{\alpha_h}{x}\right) +\sum_{a=q,\bar q}f_a(x)
D_a^h\left(\frac{\alpha_h}{x}\right)\right)
\left(\frac{x}{\alpha_h}\right)^{2\gamma}\;.
\nonumber
\eea

Now we can calculate the collinear counterterms which appear due to the
renormalization of bare PDFs and FFs. Inserting the expressions given in
Eqs.~(\ref{DGLAPpdfs}),~(\ref{DGLAPffs}) into the LO impact
factor~(\ref{LO_bis}), we obtain, after some suitable transformations
of the integration variables in the terms coming from the PDFs
renormalization, the following collinear counterterm\footnote{In principle,
one can consider different values of the factorization scale for the evolution
of PDFs and FFs. Here for simplicity we take these scales equal.}
\[
\frac{\pi\sqrt{2}\, \vec k^{\, 2}}{{\cal C}}
\frac{d\Phi^h(\nu,n)|_{\rm{coll.\ c.t.}}}
{d\alpha_h d^{2+2\epsilon}\vec k}
= -  \frac{\alpha_s}{2\pi}\left(\frac{1}{\hat \epsilon}+\ln\frac{\mu_F^2}
{\mu^2}\right)\int\limits^1_{\alpha_h} \frac{dx}{x}
\int\limits^1_{\frac{\alpha_h}{x}} \frac{dz}{z}
\left(\vec k^{\,\,2} \right)^{\gamma-{n \over 2}}
\left(\vec k \cdot \vec l \,\, \right)^n
\]
\[
\times
\left[  (1+z^{-2\gamma})P_{qq}(z)\sum_{a=q,\bar q}f_a(x)
D_a^h\left(\frac{\alpha_h}{xz}\right)+ \left(\frac{C_A}{C_F}
+z^{-2\gamma}\right) P_{gq}(z)\sum_{a=q,\bar q}f_a(x)
D_g^h\left(\frac{\alpha_h}{xz}\right)\right.
\]
\beq
\left.
+(1+z^{-2\gamma})\frac{C_A}{C_F}P_{gg}(z) f_g(x)
D_g^h\left(\frac{\alpha_h}{x z}\right)+ \frac{C_A}{C_F}
\left(\frac{C_F}{C_A}+z^{-2\gamma}\right)P_{qg}(z) f_g(x)
\sum_{a=q,\bar q}D_a^h\left(\frac{\alpha_h}{x z}\right)\right] \;.
\label{c.count.t}
\eeq

The other counterterm is related with QCD charge renormalization;
with $n_f$ active quark flavors we have
\beq
\alpha_s=\alpha_s(\mu_R)\left[1+\frac{\alpha_s(\mu_R)}{4\pi}\beta_0
\left(\frac{1}{\hat \epsilon}+\ln\frac{\mu_R^2}{\mu^2}\right)\right]
\,,\quad \beta_0=\frac{11C_A}{3}-\frac{2n_f}{3}\;,
\label{charge-ren}
\eeq
and is given by
\bea
&&
\frac{\pi\sqrt{2}\, \vec k^{\, 2}}{{\cal C}}\frac{d\Phi(\nu,n)|_{\rm{charge
\ c.t.}}}
{d\alpha d^{2+2\epsilon}\vec k}= \frac{\alpha_s}{2\pi}
\left(\frac{1}{\hat \epsilon}+\ln\frac{\mu_R^2}{\mu^2}\right)
\left(\frac{11 C_A}{6}-\frac{n_f}{3}\right)
\nonumber \\
&&
\times \int\limits^1_{\alpha_h}\frac{dx}{x}
\left( \frac{C_A}{C_F}f_g(x) D_g^h\left(\frac{\alpha_h}{x}\right)
+\sum_{a=q,\bar q}f_a(x) D_a^h\left(\frac{\alpha_h}{x}\right)\right)
\left(\vec k^{\,\,2} \right)^{\gamma-{n \over 2}}
\left(\vec k \cdot \vec l \,\, \right)^n\;.
\label{charge.count.t}
\eea

To simplify formulae, from now on we put the arbitrary scale of dimensional
regularization equal to the unity, $\mu=1$.

In what follows we will present intermediate results always for
$\frac{\pi\sqrt{2}\, \vec k^{\, 2}}{{\cal C}}\frac{d\Phi^h(\nu,n)}
{d\alpha_h d^{2+2\epsilon}\vec k}$, which we denote for shortness as
\beq
\frac{\pi\sqrt{2}\, \vec k^{\, 2}}{{\cal C}}\frac{d\Phi^h(\nu,n)}
{d\alpha_h d^{2+2\epsilon}\vec k}=I \, .
\eeq
Note that we always denote by $\vec k$ the transverse momentum of the parton
which fragments to the identified hadron. Moreover, $\alpha_s$ with no
argument from now on is to be understood as $\alpha_s(\mu_R)$. 

We will consider separately the subprocesses initiated by a quark and a
gluon PDF, and denote
\beq
I=I_q+I_g \, .
\eeq
We start with the case of incoming quark.

\subsection{Incoming quark}

We distinguish virtual corrections and real emission contributions,
\beq{}
I_q=I_q^V+I_q^R \, .
\eeq

\begin{figure}[tb]
\centering
\includegraphics[scale=0.8]{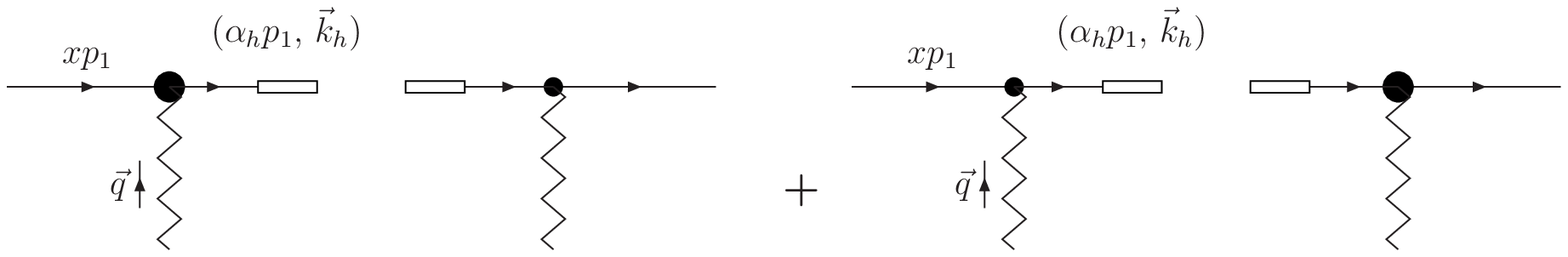}
\caption[]{Diagrammatic representation of the NLO vertex for the identified
hadron production for the case of incoming quark: virtual corrections
(the larger black blobs indicate the 1-loop contribution).}
\label{fig:vertex_quark_NLA_virt}
\end{figure}

Virtual corrections (see Fig.~\ref{fig:vertex_quark_NLA_virt}) are the same
as in the case of the inclusive quark impact factor, therefore we have
\[
I_q^V=-\frac{\alpha_s}{2\pi}\frac{\Gamma[1-\epsilon]}{(4\pi)^\epsilon}
\frac{1}{\epsilon}\frac{\Gamma^2(1+\epsilon)}{\Gamma(1+2\epsilon)}
 \int\limits^1_{\alpha_h}\frac{dx}{x}
\sum_{a=q,\bar q}f_a(x)D_a^h\left(\frac{\alpha_h}{x}\right)
\left(\vec k^{\,\,2} \right)^{\gamma+\epsilon-{n \over 2}}
\left(\vec k \cdot \vec l \,\, \right)^n
\]
\[
\times\left\{
C_F\left(\frac{2}{\epsilon}-\frac{4}{1+2\epsilon}+1\right)\right.
-n_f\frac{1+\epsilon}{(1+2\epsilon)(3+2\epsilon)}
+C_A\left(\ln\frac{s_0}{\vec k^{\,\,2}}+\psi(1-\epsilon)-2\psi(\epsilon)
+\psi(1)\right.
\]
\beq
\left.\left.
+\frac{1}{4(1+2\epsilon)(3+2\epsilon)}-\frac{2}{\epsilon(1+2\epsilon)}
-\frac{7}{4(1+2\epsilon)}-\frac{1}{2}\right)\right\}\;.
\label{Qvirt}
\eeq

We expand (\ref{Qvirt}) in $\epsilon$ and present the result as a sum of
the singular and the finite parts.   The singular contribution reads
\[
\left(I_q^V\right)_{s}=-\frac{\alpha_s}{2\pi}\frac{\Gamma[1-\epsilon]}
{(4\pi)^\epsilon}\frac{1}{\epsilon}
\frac{\Gamma^2(1+\epsilon)}{\Gamma(1+2\epsilon)}  \int\limits^1_{\alpha_h}
\frac{dx}{x}\sum_{a=q,\bar q}f_a(x) D_a^h\left(\frac{\alpha_h}{x}
\right)\left(\vec k^{\,\,2}\right)^{\gamma+\epsilon-{n \over 2}}
\left(\vec k \cdot \vec l \,\, \right)^n
\]
\beq
\times \left\{C_F\left(\frac{2}{\epsilon}-3\right)-\frac{n_f}{3}
+C_A\left(\ln\frac{s_0}{\vec k^2}+\frac{11}{6}\right)
\right\}\;,
\label{Qvirt-div}
\eeq
whereas for the regular part we obtain
\[
\left(I_q^V\right)_{r}=-\frac{\alpha_s}{2\pi}
 \int\limits^1_{\alpha_h}\frac{dx}{x}
\sum_{a=q,\bar q}f_a(x) D_a^h
\left(\frac{\alpha_h}{x}\right)\left(\vec k^{\,\,2} 
\right)^{\gamma-{n \over 2}}
\left(\vec k \cdot \vec l \,\, \right)^n
\]
\beq
\times \left\{8 C_F+\frac{5n_f}{9}
-C_A\left(\frac{85}{18}+\frac{\pi^2}{2}\right)\right\}\;.
\label{Qvirt-fin}
\eeq
Note that $\left(I_q^V\right)_{s}+\left(I_q^V\right)_{r}$ differs from $I_q^V$
by terms which are ${\cal O}(\epsilon)$.

\subsubsection{Quark-gluon intermediate state}
\label{subsub_QG}

The starting point here is the quark-gluon intermediate state contribution
to the inclusive quark impact factor,
\beq
\Phi^{\{QG\}}=\Phi_q g^2\vec q^{\,\, 2}\frac{d^{2+2\epsilon} \vec k_1}
{(2\pi)^{3+2\epsilon}}\frac{d\beta_1}{\beta_1}\frac{[1+\beta_2^2
+\epsilon \beta_1^2]}{\vec k_1^{\, 2} \vec k_2^{\, 2}
(\vec k_2\beta_1-\vec k_1 \beta_2)^2} \left\{C_F \beta_1^2\vec k_2^{\, 2}
+C_A\beta_2\left(\vec k_1^{\,2}-\beta_1\vec k_1\cdot\vec q\right) \right\}\;,
\eeq
where $\beta_1$ and $\beta_2$ are the relative longitudinal momenta
($\beta_1+\beta_2=1$) and $\vec k_1$ and $\vec k_2$ are the transverse
momenta ($\vec k_1+\vec k_2=\vec q$) of the produced gluon and quark,
respectively.

\begin{figure}[tb]
\centering
\includegraphics[scale=0.8]{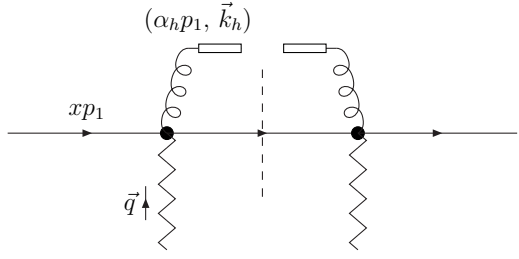}
\caption[]{Diagrammatic representation of the NLO vertex for the identified
hadron production for the case of incoming quark: real corrections
from quark-gluon intermediate state, case of gluon fragmentation.}
\label{fig:vertex_quark_NLA_gfrag}
\end{figure}

We need to consider separately the fragmentation of the quark and of the
gluon into the final hadron, therefore  we denote the corresponding
contributions as $I_{q,q}^R$ and $I_{q,g}^R$,
\beq{}
I_q^R=I_{q,q}^R+I_{q,g}^R \; .
\eeq
We start with the gluon fragmentation case, $I_{q,g}^R$ (see
Fig.~\ref{fig:vertex_quark_NLA_gfrag}).

{\bf a) gluon fragmentation}

The ``parent parton'' variables are $\vec k=\vec k_1$, $\zeta=\beta_1$
($\beta_2=\bar \zeta\equiv 1-\zeta$, $\vec k_2=\vec q-\vec k$), therefore we 
have
\[
I_{q,g}^R=\frac{\alpha_s}{2\pi(4\pi)^\epsilon}
\int\frac{ d^{2+2\epsilon}\vec q}{\pi^{1+\epsilon}}\left(\vec q^{\,\,2} \right)
^{\gamma-{n \over 2}}
\left(\vec q \cdot \vec l \,\, \right)^n \int\limits^1_{\alpha_h} \frac{dx}{x}
\int\limits^1_{\frac{\alpha_h}{x}} \frac{d\zeta}{\zeta}\sum_{a=q,\bar q}
f_a(x)D^h_g\left(\frac{\alpha_h}{x \zeta}\right)
\]
\beq
\times \frac{1+\bar \zeta^2 +\epsilon \zeta^2}{\zeta}
\left[
C_F\frac{1}{\left(\vec q-\frac{\vec k}{\zeta}\right)^2}
+C_A\frac{\bar \zeta}{\zeta}
\frac{\frac{\vec k^{\,\,2}}{\zeta}-\vec k \cdot \vec q}
{(\vec q-\vec k)^2\left(\vec q-\frac{\vec k}{\zeta}\right)^2}\right]\;.
\label{QGgluon}
\eeq

It is worth stressing the difference between the previous calculations of
NLO inclusive parton impact factors and the present case of fragmentation
to a hadron with fixed momentum. In  parton impact factor case, one keeps
fixed the Reggeon transverse momentum $\vec q$ and integrates over the
allowed phase space of the produced partons, i.e. the integration is
of the form $\int \frac{d\zeta}{2\zeta(1-\zeta)}d^{2+2\epsilon}\vec k\dots$
In the fragmentation case, instead, we keep fixed the
momentum of the parent parton $\zeta, \vec k$, and allow the Reggeon momentum
$\vec q$ to vary. Indeed, the expression~(\ref{QGgluon}) contains the explicit
integration over the momentum $\vec q$ with the LO BFKL eigenfunctions,
which is needed in order to obtain the impact factor in the
$(\nu,n)$-representation.

The $\vec q$-integration in (\ref{QGgluon}) generates $1/\epsilon$ poles due
to the integrand singularities at $\vec q\to \vec k/\zeta$ for the
contribution proportional to $C_F$ and at $\vec q\to \vec k$ for the one
proportional to $C_A$.
Accordingly we split the result of the $\vec q$-integration into the sum of
two terms: ``singular'' and ``non-singular'' parts. The non-singular part is
defined as
\[
\frac{\alpha_s}{2\pi(4\pi)^\epsilon}
 \int\limits^1_{\alpha_h} \frac{dx}{x} \int\limits^1_{\frac{\alpha_h}{x}}
\frac{d\zeta}{\zeta}\sum_{a=q,\bar q} f_a(x)
D^h_g\left(\frac{\alpha_h}{x \zeta}\right)C_A
\frac{\bar \zeta}{\zeta}\left(\frac{1+\bar \zeta^2 +\epsilon \zeta^2}{\zeta}
\right)
\]
\[
\times \int \frac{d^{2+2\epsilon}\vec q}{\pi^{1+\epsilon}}
\frac{\frac{\vec k^{\,\,2}}{\zeta}-\vec k \cdot \vec q}{(\vec q-\vec k)^2
\left(\vec q-\frac{\vec k}{\zeta}\right)^2}
\left[\left(\vec q^{\,\,2} \right)^{\gamma-{n \over 2}}
\left(\vec q \cdot \vec l \,\, \right)^n
-\left(\vec k^{\,\,2} \right)^{\gamma-{n \over 2}}
\left(\vec k \cdot \vec l \,\, \right)^n\right]
\]
\[
= \frac{\alpha_s}{2\pi(4\pi)^\epsilon}
 \int\limits^1_{\alpha_h} \frac{dx}{x} \int\limits^1_{\frac{\alpha_h}{x}}
\frac{d\zeta}{\zeta}\sum_{a=q,\bar q} f_a(x)
D^h_g\left(\frac{\alpha_h}{x \zeta}\right)\left(\vec k^{\,\,2} 
\right)^{\gamma+\epsilon-{n \over 2}}\left(\vec k \cdot
\vec l \,\, \right)^n C_A
\frac{\bar \zeta}{\zeta}\left(\frac{1+\bar \zeta^2 +\epsilon \zeta^2}{\zeta}
\right)
\]
\beq
\times \int \frac{d^{2+2\epsilon}\vec a}{\pi^{1+\epsilon}}
\frac{\frac{1}{\zeta}-\vec n \cdot \vec a}{(\vec a-\vec n)^2\left(\vec a
-\frac{\vec n}{\zeta}\right)^2}\left[\left(\vec a^{\,\,2} \right)^{\gamma
-{n \over 2}}\left(\frac{\vec a \cdot \vec l }{\vec n \cdot \vec l \,\, }
\right)^n-1\right]\; ,
\label{QGgluon-reg}
\eeq
where $\vec n$ is a unit vector, $\vec n^{\, 2}=1$. Taking this expression for 
$\epsilon=0$ we have
\beq{}
\left(I_{q,g}^R\right)_r=\frac{\alpha_s}{2\pi}
\int\limits^1_{\alpha_h} \frac{dx}{x} \int\limits^1_{\frac{\alpha_h}{x}}
\frac{d\zeta}{\zeta}\sum_{a=q,\bar q} f_a(x)
D^h_g\left(\frac{\alpha_h}{x \zeta}\right)
\left(\vec k^{\,\,2} \right)^{\gamma-{n \over 2}}\left(\vec k \cdot
\vec l \,\, \right)^n
C_A
\frac{\bar \zeta}{\zeta}\left(\frac{1+\bar \zeta^2}{\zeta}
\right)I_1 \;,
\eeq
where we define the function
\beq{}
I_1= I_1(n,\gamma,\zeta)=\int \frac{d^{2}\vec a}{\pi}
\frac{\frac{1}{\zeta}-\vec n \cdot \vec a}{(\vec a-\vec n)^2\left(\vec a
-\frac{\vec n}{\zeta}\right)^2}\left[\left(\vec a^{\,\,2} \right)^{\gamma
} e^{i n\phi}-1\right]\;,
\eeq
with the azimuth of the vector $\vec a$ ($\phi$) counted from the
direction of the unit vector $\vec n$.

For the singular contribution we obtain
\[
\frac{\alpha_s}{2\pi}\frac{\Gamma[1-\epsilon]}{(4\pi)^\epsilon}\frac{1}
{\epsilon}\frac{\Gamma^2(1+\epsilon)}{\Gamma(1+2\epsilon)}
\int\limits^1_{\alpha_h} \frac{dx}{x}\int\limits^1_{\frac{\alpha_h}{x}}
\frac{d\zeta}{\zeta}
\sum_{a=q,\bar q} f_a(x) D^h_g\left(\frac{\alpha_h}{x \zeta}\right)
\left(\vec k^{\,\,2} \right)^{\gamma+\epsilon-{n \over 2}}
\left(\vec k \cdot \vec l \,\, \right)^n
\]
\[
\times \frac{1+\bar \zeta^2+\epsilon\zeta^2}{\zeta}\left[
C_F\frac{\Gamma(1+2\epsilon)\Gamma(\frac{n}{2}-\gamma-\epsilon)
\Gamma(\frac{n}{2}+1+\gamma+\epsilon)}
{\Gamma(1+\epsilon)\Gamma(1-\epsilon)\Gamma(\frac{n}{2}-\gamma)
\Gamma(\frac{n}{2}+1+\gamma+2\epsilon)}
\zeta^{-2\epsilon-2\gamma}+C_A\left(\frac{\bar \zeta}{\zeta}
\right)^{2\epsilon}\right] \;.
\]

Expanding it in $\epsilon$ we get
\bea
&&
\left(I_{q,g}^R\right)_s=\frac{\alpha_s}{2\pi}\frac{\Gamma[1-\epsilon]}
{\epsilon (4\pi)^\epsilon}
\int\limits^1_{\alpha_h} \frac{dx}{x} \int\limits^1_{\frac{\alpha_h}{x}}
\frac{d\zeta}{\zeta} \sum_{a=q,\bar q}f_a(x)
D^h_g\left(\frac{\alpha_h}{x \zeta}\right)
\left(\vec k^{\,\,2} \right)^{\gamma+\epsilon-{n \over 2}}
\left(\vec k \cdot \vec l \,\, \right)^n \label{QGgluon-div} \\
&&
\times \left\{ P_{gq}(\zeta)\left[\frac{C_A}{C_F}
+\zeta^{-2\gamma}\right]  \right. \nonumber \\
&&
\left.
+\, \epsilon
\left(
\frac{1+\bar \zeta^2}{\zeta}\left[C_F\zeta^{-2\gamma}(\chi(n,\gamma)-2\ln\zeta)
+2C_A\ln\frac{\bar \zeta}{\zeta}\right]
+\zeta(C_F\zeta^{-2\gamma}+C_A)\right) \right\}
\;, \nonumber
\eea
where
\beq
\chi(n,\gamma)=2\psi(1)-\psi\left(\frac{n}{2}-\gamma\right)-\psi\left(\frac{n}
{2}+1+\gamma\right)
\label{bfkl-eigenvalue}
\eeq
is the eigenvalue of the LO BFKL kernel, up to the factor $N\alpha_s/\pi$.

Note that the divergent part of~(\ref{QGgluon-div}) is canceled  by 
the corresponding term of the collinear counterterm~(\ref{c.count.t}).

\begin{figure}[tb]
\centering
\includegraphics[scale=0.8]{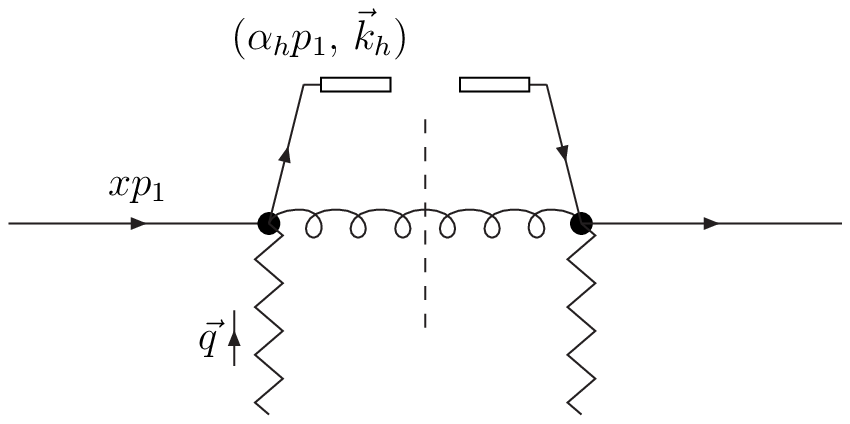}
\caption[]{Diagrammatic representation of the NLO vertex for the identified
hadron production for the case of incoming quark: real corrections
from quark-gluon intermediate state, case of quark fragmentation.}
\label{fig:vertex_quark_NLA_qfrag}
\end{figure}

{\bf b) quark fragmentation}

Now the ``parent parton'' variables are $\vec k=\vec k_2$, $\zeta=\beta_2$
($\beta_1=\bar \zeta$, $\vec k_1=\vec q-\vec k$)
(see Fig.~\ref{fig:vertex_quark_NLA_qfrag}). The corresponding
contribution reads
\[
I_{q,q}^R=\frac{\alpha_s}{2\pi(4\pi)^\epsilon}
\int \frac{d^{2+2\epsilon}\vec q}{\pi^{1+\epsilon}}\left(\vec q^{\,\,2}
\right)^{\gamma-{n \over 2}}
\left(\vec q \cdot \vec l \,\, \right)^n \int\limits^1_{\alpha_h}\frac{dx}{x}
\int\limits^1_{\frac{\alpha_h}{x}} \frac{d\zeta}{\zeta}
 \sum_{a=q,\bar q} f_a(x)D^h_a\left(\frac{\alpha_h}{x \zeta}\right)
\]
\beq
\times\frac{1+ \zeta^2 +\epsilon \bar \zeta^2}{(1- \zeta)}
\left[
C_F\frac{\bar \zeta^2}{\zeta^2}\frac{\vec k^{\, 2}}{(\vec q-\vec k)^2
\left(\vec q-\frac{\vec k}{\zeta}\right)^2}
+C_A\frac{\vec q^{\,\,2}-\vec k \cdot \vec q\, \frac{1+\zeta}{\zeta}
+\frac{\vec k^{\,2}}{\zeta}}{(\vec q-\vec k\,)^2
\left(\vec q-\frac{\vec k}{\zeta}\right)^2}
\right] \;.
\label{QGquark}
\eeq

We will consider separately the contributions proportional to $C_F$ and $C_A$.

{\bf b$_1$) quark fragmentation: $C_F$-term}

Note that the integrand of the $C_F$-term is not singular at $\zeta\to 1$.
We use the decomposition
\[
\frac{\vec k^{\, 2}}{(\vec q-\vec k)^2\left(\vec q-\frac{\vec k}{\zeta}
\right)^2} = \frac{\vec k^{\, 2}}{(\vec q-\vec k)^2+\left(\vec q
-\frac{\vec k}{\zeta}\right)^2}\left(\frac{1}{(\vec q-\vec k)^2}+
\frac{1}{\left(\vec q-\frac{\vec k}{\zeta}\right)^2}\right)
\]
in order to separate the regular and singular contributions. The regular part
is given by
\[
\frac{\alpha_s}{2\pi(4\pi)^\epsilon}\int\limits^1_{\alpha_h} \frac{dx}{x}
\int\limits^1_{\frac{\alpha_h}{x}} \frac{d\zeta}{\zeta}\sum_{a=q,\bar q}
f_a(x)D^h_a\left(\frac{\alpha_h}{x \zeta}\right)
\left(\vec k^{\,\,2} \right)^{\gamma+\epsilon-{n \over 2}}
\left(\vec k \cdot \vec l \,\, \right)^n
\frac{\bar \zeta^2}{\zeta^2}\left(\frac{1+ \zeta^2 +\epsilon \bar \zeta^2}
{1-\zeta}\right)
\]
\beq
\times C_F\int \frac{d^{2+2\epsilon}\vec a}{\pi^{1+\epsilon}}
\frac{1}{(\vec a-\vec n)^2+\left(\vec a-\frac{\vec n}{\zeta}\right)^2}
 \left[\frac{\left(\vec a^{\,\,2} \right)^{\gamma-{n \over 2}}
\left(\frac{\vec a \cdot \vec l }{\vec n \cdot \vec l \,\, }\right)^n-1}
{(\vec a-\vec n)^2}+\frac{\left(\vec a^{\,\,2} \right)^{\gamma-{n \over 2}}
\left(\frac{\vec a \cdot \vec l }{\vec n \cdot \vec l \,\, }\right)^n
-\zeta^{-2\gamma}}{(\vec a-\frac{\vec n}{\zeta})^2} \right]\;.
\label{QGquark-CF-reg}
\eeq
Therefore for $\epsilon=0$ we have
\beq{}
\left(I_{q,q}^R\right)_r^{C_F}=\frac{\alpha_s}{2\pi}\int\limits^1_{\alpha_h} 
\frac{dx}{x}
\int\limits^1_{\frac{\alpha_h}{x}} \frac{d\zeta}{\zeta}\sum_{a=q,\bar q}
f_a(x)D^h_a\left(\frac{\alpha_h}{x \zeta}\right)
\left(\vec k^{\,\,2} \right)^{\gamma-{n \over 2}}
\left(\vec k \cdot \vec l \,\, \right)^n
\frac{\bar \zeta (1+\zeta^2)}{\zeta^2}\, C_F \, I_2\ ,
\eeq
where we define the function
\beq{}
  I_2=I_2(n,\gamma,\zeta)= \int \frac{d^{2}\vec a}{\pi}
\frac{1}{(\vec a-\vec n)^2+\left(\vec a-\frac{\vec n}{\zeta}\right)^2}
 \left[\frac{\left(\vec a^{\,\,2} \right)^{\gamma}e^{i n\phi}-1}
{(\vec a-\vec n)^2}+\frac{\left(\vec a^{\,\,2} \right)^{\gamma}e^{i n\phi}
-\zeta^{-2\gamma}}{(\vec a-\frac{\vec n}{\zeta})^2} \right]\;.
\eeq

The singular part is proportional to the integral
\[
\int \frac{d^{2+2\epsilon}\vec q}{\pi^{1+\epsilon}} \frac{\vec k^{\, 2}}
{(\vec q-\vec k)^2+\left(\vec q-\frac{\vec k}{\zeta}\right)^2}
\left[
\frac{\left(\vec k^{\,\,2} \right)^{\gamma-{n \over 2}}\left(\vec k \cdot
\vec l \,\, \right)^n}{(\vec q-\vec k)^2}+
\frac{\left(\left({\vec k\over \zeta }\right)^{\,\,2} \right)^{\gamma
-{n \over 2}}\left({\vec k \over \zeta} \cdot \vec l \,\, \right)^n}
{(\vec q-{\vec k\over \zeta})^2}\right]
\]
\[
=\frac{\Gamma(1-\epsilon)\Gamma^2(1+\epsilon)}{\epsilon\Gamma[1+2\epsilon]}
\left(\vec k^{\,\,2} \right)^{\gamma+\epsilon-{n \over 2}}
\left(\vec k \cdot \vec l \,\, \right)^n \left(\frac{\bar \zeta}{\zeta}
\right)^{2\epsilon-2}\left(1+\zeta^{-2\gamma}\right)\;,
\]
therefore, for the singular part of the $C_F$-term we have
\[
\frac{\alpha_s}{2\pi}\frac{\Gamma(1-\epsilon)}{\epsilon (4\pi)^\epsilon}
\frac{\Gamma^2(1+\epsilon)}{\Gamma(1+2\epsilon)}
\int\limits^1_{\alpha_h}
\frac{dx}{x}\int\limits^1_{\frac{\alpha_h}{x}} \frac{d\zeta}{\zeta }
 \sum_{a=q,\bar q}f_a(x)
D_a^h\left(\frac{\alpha_h}{x\zeta }\right)
\left(\vec k^{\,\,2} \right)^{\gamma+\epsilon-{n \over 2}}
\left(\vec k \cdot \vec l \,\, \right)^n
\]
\beq
\times C_F\frac{1+ \zeta^2 +\epsilon \bar \zeta^2}{1- \zeta}
\left(\frac{\bar \zeta}{\zeta}\right)^{2\epsilon}\left(1+\zeta^{-2\gamma}
\right)\;.
\label{QGquark-CF-sing}
\eeq
The next step is to introduce the plus-prescription, which is defined as
\beq
\int\limits^1_a d \zeta \frac{F(\zeta)}{(1-\zeta)_+}
=\int\limits^1_a d \zeta \frac{F(\zeta)-F(1)}{(1-\zeta)}
-\int\limits^a_0 d \zeta \frac{F(1)}{(1-\zeta)}\; ,
\label{plus}
\eeq
for any function $F(\zeta)$, regular at $\zeta=1$. Note that
\[
(1-\zeta)^{2\epsilon-1}=(1-\zeta)^{2\epsilon -1}_+ +\frac{1}{2\epsilon}
\delta(1-\zeta)=\frac{1}{2\epsilon}\delta(1-\zeta)+\frac{1}{(1-\zeta)_+}
+2\epsilon\left(\frac{\ln (1-\zeta)}{1-\zeta}\right)_+ +{\cal O}(\epsilon^2)
\; .
\]
Using this result, one can write
\[
C_F \frac{1+ \zeta^2 +\epsilon \bar \zeta^2}{1-\zeta}
\left(\frac{\bar \zeta}{\zeta}\right)^{2\epsilon}\left(1+\zeta^{-2\gamma}
\right)= C_F\left[\frac{2}{\epsilon}\delta(1-\zeta)+ \frac{1+\zeta^2}
{(1-\zeta)_+}\left(1+\zeta^{-2\gamma}\right) +{\cal O}(\epsilon )\right]
\]
\[
= C_F\left[\left(\frac{2}{\epsilon}-3\right)\delta(1-\zeta)
+\left( \frac{1+\zeta^2}{(1-\zeta)_+}+ \frac{3}{2}\delta(1-\zeta)\right)
\left(1+\zeta^{-2\gamma}\right) +{\cal O}(\epsilon )\right]
\]
\[
= C_F\left(\frac{2}{\epsilon}-3\right)\delta(1-\zeta)
+P_{qq}(\zeta)\left(1+\zeta^{-2\gamma}\right) +{\cal O}(\epsilon ) \;.
\]
Taking this into account and expanding in $\epsilon$ the singular part
of the $C_F$-term, one gets the following result for the divergent
contribution:
\bea
&&
\left(I_{q,q}^R\right)_s^{C_F}=
\frac{\alpha_s}{2\pi}\frac{\Gamma[1-\epsilon]}{\epsilon (4\pi)^\epsilon}
\frac{\Gamma^2(1+\epsilon)}{\Gamma(1+2\epsilon)}
\int\limits^1_{\alpha_h} \frac{dx}{x} \int\limits^1_{\frac{\alpha_h}{x}}
\frac{d\zeta}{\zeta} \sum_{a=q,\bar q}f_a(x)
D^h_a\left(\frac{\alpha_h}{x \zeta}\right)
\left(\vec k^{\,\,2} \right)^{\gamma+\epsilon-{n \over 2}}
\left(\vec k \cdot \vec l \,\, \right)^n
\nonumber \\
&&
\times
\left\{C_F \left(\frac{2}{\epsilon}-3\right)\delta(1-\zeta)
+P_{qq}(\zeta)\left(1+\zeta^{-2\gamma}\right) \right. \nonumber \\
&&
\left.
+ \, \epsilon \, C_F \, (1+\zeta^{-2\gamma})\left(\bar \zeta+2(1+\zeta^2)
\left(\frac{\ln (1-\zeta)}{1-\zeta}\right)_+ -2(1+ \zeta^2)\frac{\ln \zeta}
{1-\zeta}\right)
\right\} \;.
\label{QGquark-CF-div}
\eea

Note that the first term of the divergent contribution given
in~(\ref{QGquark-CF-div}) cancels in the sum with the corresponding term of
the virtual correction~(\ref{Qvirt-div}), whereas the second one is canceled
by the corresponding term of the collinear counterterm~(\ref{c.count.t}).

{\bf b$_2$) quark fragmentation: $C_A$-term}

The $C_A$-contribution needs a special treatment due to the behavior
of~(\ref{QGquark}) in the region $\zeta\to 1$.

We use the following decomposition:
\[
\frac{C_A}{(1- \zeta)}
\left(\frac{\vec q^{\,\,2}-\vec k \cdot \vec q\, \frac{1+\zeta}{\zeta}
+\frac{\vec k^{\,\,2}}{\zeta}}{(\vec q-\vec k\,)^2\left(\vec q-\frac{\vec k}
{\zeta}\right)^2}\right)=\frac{C_A}{2}\frac{2}{(\vec q-\vec k)^2}\frac{1}
{(1- \zeta )}
\]
\[
+ \frac{C_A}{2(1-\zeta)}\left[\frac{1}{(\vec q-\frac{\vec k}{\zeta})^2}
-\frac{1}{(\vec q-\vec k)^2}-\left(\frac{\bar \zeta}{\zeta}\right)^2
\frac{\vec k^2}{(\vec q-\vec k)^2(\vec q-\frac{\vec k}{\zeta})^2}\right] \;.
\]
The second term in the r.h.s. is regular for $\zeta\to 1$ and can be treated
similarly to what we did above in the case of the $C_F$-contribution.
The first term is singular and the integration over $\zeta$ has to be
restricted (according to definition of NLO impact factor) by the requirement
\[
M^2_{QG}\leq s_\Lambda \;,\;\;\;\;
M^2_{QG}=\frac{\vec k_1^2}{\beta_1}+\frac{\vec k_2^2}{\beta_2}-\vec q^{\,\, 2}
=\frac{(\vec q-\vec k)^2}{1-\zeta}+\frac{\vec k^2}{\zeta}-\vec q^{\,\, 2}\; ,
\]
and assuming $s_\Lambda\to \infty$. Therefore the $\zeta$-integral has the form
\[
\int\limits^{1-\zeta_0}_{a} d \zeta \frac{F(\zeta)}{1-\zeta}\;,\;\;\;\;
{\rm for} \;\;\;\;\; \zeta_0=\frac{(\vec q-\vec k)^2}{s_\Lambda}\to 0\; .
\]
Using the plus-prescription~(\ref{plus}) one can write
\beq
\int\limits^{1-\zeta_0}_{a} d \zeta \frac{F(\zeta)}{1-\zeta}
=\int\limits^1_a d \zeta \frac{F(\zeta)}{(1-\zeta)_+}
+ F(1)\ln\frac{1}{\zeta_0} \;, \;\;\;\; {\rm for } \;\; \zeta_0\to 0\;,
\label{zeta0}
\eeq
for any function not singular in the limit $\zeta\to 1$, and
\[
\frac{C_A}{(1- \zeta)}
\left(\frac{\vec q^{\,\,2}-\vec k \cdot \vec q\, \frac{1+\zeta}{\zeta}
+\frac{\vec k^{\,\,2}}{\zeta}}{(\vec q-\vec k\,)^2\left(\vec q-\frac{\vec k}
{\zeta}\right)^2}\right)=\frac{C_A}{2}\delta(1-\zeta)\frac{2}
{(\vec q-\vec k)^2}\ln\frac{s_\Lambda}{(\vec q-\vec k)^2 }
\]
\[
+ \frac{C_A}{2}\frac{2}{(\vec q-\vec k)^2}\frac{1}{(1- \zeta)_+ }+
\frac{C_A}{2(1-\zeta)}\left[\frac{1}{(\vec q-\frac{\vec k}{\zeta})^2}
-\frac{1}{(\vec q-\vec k)^2}
-\left(\frac{\bar \zeta}{\zeta}\right)^2\frac{\vec k^2}{(\vec q-\vec k)^2
(\vec q-\frac{\vec k}{\zeta})^2}\right] \;.
\]
We remind that the definition of NLO impact factor requires the
subtraction of the contribution coming from the gluon emission in the central
rapidity region, given by the last term in Eq.~(\ref{eq:a19}), which we call
below ``BFKL subtraction term''. After this subtraction the parameter
$s_\Lambda$ should be sent to infinity, $s_\Lambda\to \infty$. Our simple
treatment of the invariant mass constraint, $M^2_{QG}\leq s_\Lambda$,
anticipates this limit $s_\Lambda\to \infty$, therefore we neglect all
contributions which are suppressed by powers of $1/s_\Lambda$. Moreover,
the first term in the r.h.s. of the above equation should be naturally
combined with the BFKL subtraction term, giving finally
\[
\frac{C_A}{(1- \zeta)}\left(\frac{\vec q^{\,\,2}-\vec k \cdot \vec q\,
\frac{1+\zeta}{\zeta}+\frac{\vec k^{\,\,2}}{\zeta}}{(\vec q-\vec k\,)^2
\left(\vec q-\frac{\vec k}{\zeta}\right)^2}\right)\to\frac{C_A}{2}
\delta(1-\zeta)\frac{1}{(\vec q-\vec k)^2}\ln\frac{s_0}{(\vec q-\vec k)^2 }
\]
\[
+\frac{C_A}{2}\frac{2}{(\vec q-\vec k)^2}\frac{1}{(1- \zeta)_+ }+
\frac{C_A}{2(1-\zeta)}\left[\frac{1}{(\vec q-\frac{\vec k}{\zeta})^2}
-\frac{1}{(\vec q-\vec k)^2}
-\left(\frac{\bar \zeta}{\zeta}\right)^2\frac{\vec k^2}{(\vec q-\vec k)^2
(\vec q-\frac{\vec k}{\zeta})^2}\right]\;.
\]
After that, we are ready to perform the $\vec q$-integration, which naturally
introduces the separation into singular and non-singular contributions. The
singular contribution reads
\[
\frac{\alpha_s}{2\pi}\frac{\Gamma(1-\epsilon)}{\epsilon (4\pi)^\epsilon}
\frac{\Gamma^2(1+\epsilon)}{\Gamma(1+2\epsilon)}
\int\limits^1_{\alpha_h}
\frac{dx}{x}\int\limits^1_{\frac{\alpha_h}{x}} \frac{d\zeta}{\zeta }
 \sum_{a=q,\bar q}f_a(x)
D_a^h\left(\frac{\alpha_h}{x\zeta }\right)
\left(\vec k^{\,\,2} \right)^{\gamma+\epsilon-{n \over 2}}
\left(\vec k \cdot \vec l \,\, \right)^n
\]
\[
\times \frac{C_A}{2}\left(1+ \zeta^2 +\epsilon \bar \zeta^2\right)
\left\{
\frac{\Gamma(1+2\epsilon)\Gamma(\frac{n}{2}-\gamma-\epsilon)\Gamma(\frac{n}{2}
+1+\gamma+\epsilon)}{\Gamma(1+\epsilon)\Gamma(1-\epsilon)
\Gamma(\frac{n}{2}-\gamma)\Gamma(\frac{n}{2}+1+\gamma+2\epsilon)}
\right.
\]
\[
\times
\left[
\delta(1-\zeta)\left(\ln\frac{s_0}{\vec k^2}+\psi\biggl(\frac{n}{2}-\gamma
-\epsilon\biggr)+\psi\biggl(1+\gamma+\frac{n}{2}+2\epsilon\biggr)
-\psi(\epsilon)-\psi(1)\right)+\frac{2}{(1-\zeta)_+}\right.
\]
\beq
\left.\left.
+\frac{(\zeta^{-2\epsilon-2\gamma}-1)}{1-\zeta}\right]-\bar \zeta^{2\epsilon-1}
\left(\zeta^{-2\epsilon}+\zeta^{-2\gamma-2\epsilon}\right)
\right\}\;.
\label{QGquark-CA-sing}
\eeq
Expanding this expression in $\epsilon$ and using that
\[
\frac{(\zeta^{-2\epsilon-2\gamma}-1)}{1-\zeta}=
\frac{(\zeta^{-2\epsilon-2\gamma}-1)}{(1-\zeta)_+} \;,
\]
and
\[
\bar \zeta^{2\epsilon-1}\left(\zeta^{-2\epsilon}+\zeta^{-2\gamma-2\epsilon}
\right)=
\left(\zeta^{-2\epsilon}+\zeta^{-2\gamma-2\epsilon}\right)
\left(\frac{\delta(1-\zeta)}{2\epsilon}+\frac{1}{(1-\zeta)_+}
+{\cal O}(\epsilon)\right)\;,
\]
we get a divergent term,
\bea
&&
\left(I_{q,q}^R\right)_s^{C_A}=\frac{\alpha_s}{2\pi}\frac{\Gamma(1-\epsilon)}
{\epsilon (4\pi)^\epsilon}
\int\limits^1_{\alpha_h} \frac{dx}{x}\int\limits^1_{\frac{\alpha_h}{x}}
\frac{d\zeta}{\zeta } \sum_{a=q,\bar q}
f_a(x) D_a^h\left(\frac{\alpha_h}{x\zeta }\right)
\left(\vec k^{\,\,2} \right)^{\gamma+\epsilon
-{n \over 2}}\left(\vec k \cdot \vec l \,\, \right)^n
\nonumber \\
&&
\times  \left\{C_A \, \delta(1-\zeta)\ln\frac{s_0}{\vec k^2}\right.
\label{QGquark-CA-div}
\\
&&
+\, \epsilon \, C_A \, \left[\delta(1-\zeta)\left(\chi(n,\gamma)\ln\frac{s_0}
{\vec k^2}
+\frac{1}{2}\left(\psi^\prime\left(1+\gamma+\frac{n}{2}\right)
-\psi^\prime\left(\frac{n}{2}-\gamma\right)-\chi^2(n,\gamma)\right)\right)
\right.
\nonumber \\
&&
\left.\left.
+(1+\zeta^2)\left((1+\zeta^{-2\gamma})\left(\frac{\chi(n,\gamma)}
{2(1-\zeta)_+}-\left(\frac{\ln (1-\zeta)}{1-\zeta}\right)_+\right)
+\frac{\ln\zeta}{1-\zeta}\right)\right]\right\}\;. \nonumber
\eea

The divergent term given in~(\ref{QGquark-CA-div}) cancels in the sum
with the corresponding term of the virtual correction~(\ref{Qvirt-div}). The
remaining singularity in~(\ref{Qvirt-div}) vanishes after the charge
renormalization~(\ref{charge-ren}).

The regular contribution differs from~(\ref{QGquark-CF-reg}) only by one
factor and reads
\[
\left(I_{q,q}^R\right)_r^{C_A}=
\frac{\alpha_s}{2\pi}
\int\limits^1_{\alpha_h} \frac{dx}{x} \int\limits^1_{\frac{\alpha_h}{x}}
\frac{d\zeta}{\zeta}\sum_{a=q,\bar q} f_a(x)
D^h_a\left(\frac{\alpha_h}{x \zeta}\right)
\left(\vec k^{\,\,2} \right)^{\gamma-{n \over 2}}\left(\vec k \cdot
\vec l \,\, \right)^n
\frac{\bar \zeta}{\zeta^2}\left(1+ \zeta^2\right)
\]
\beq
\times
\left(-\frac{C_A}{2}\right)I_2\; .
\label{QGquark-CA-reg}
\eeq

\begin{figure}[tb]
\centering
\includegraphics[scale=0.8]{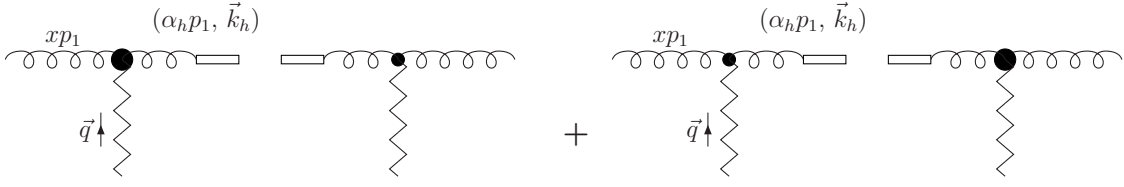}
\caption[]{Diagrammatic representation of the NLO vertex for the identified
hadron production for the case of incoming gluon: virtual corrections
(the larger black blobs indicate the 1-loop contribution).}
\label{fig:vertex_gluon_NLA_virt}
\end{figure}

\subsection{Incoming gluon}

We distinguish virtual corrections and real emission contributions,
\beq{}
I_g=I_g^V+I_g^R \; .
\eeq

Virtual corrections (see Fig.~\ref{fig:vertex_gluon_NLA_virt}) are the same
as in the case of inclusive gluon impact factor,
\[
I_g^V=
-\frac{\alpha_s}{2\pi}\frac{\Gamma[1-\epsilon]}{(4\pi)^\epsilon}\frac{1}
{\epsilon}\frac{\Gamma^2(1+\epsilon)}{\Gamma(1+2\epsilon)}
 \int\limits^1_{\alpha_h}\frac{dx}{x}
f_g(x) D_{g}^h\left(\frac{\alpha_h}{x}\right)\left(\vec k^{\,\,2} 
\right)^{\gamma+\epsilon-{n \over 2}}
\left(\vec k \cdot \vec l \,\, \right)^n \, \frac{C_A}{C_F}
\]
\[
\times \biggl\{\biggr.
C_A\left(\ln\frac{s_0}{\vec k^{\,\,2}}+\frac{2}{\epsilon}-\frac{11+9\epsilon}
{2(1+2\epsilon)(3+2\epsilon)}+\psi(1-\epsilon)-2\psi(1+\epsilon)+\psi(1)
\right.
\]
\beq
\left.\left.
+\frac{\epsilon}{(1+\epsilon)(1+2\epsilon)(3+2\epsilon)}\right)
+n_f\left(\frac{(1+\epsilon)(2+\epsilon)-1-\frac{\epsilon}{1+\epsilon}}
{(1+\epsilon)(1+2\epsilon)(3+2\epsilon)}\right)\right\}\;.
\label{Gvirt}
\eeq

Expanding it in $\epsilon$ we obtain the following results for the singular,
\[
\left(I_g^V\right)_s=-\frac{\alpha_s}{2\pi}\frac{\Gamma[1-\epsilon]}
{(4\pi)^\epsilon}\frac{1}{\epsilon}\frac{\Gamma^2(1+\epsilon)}
{\Gamma(1+2\epsilon)} \int\limits^1_{\alpha_h}\frac{dx}{x}
f_g(x) D_g^h\left(\frac{\alpha_h}{x}\right)\left(\vec k^{\,\,2} 
\right)^{\gamma+\epsilon-{n \over 2}}
\left(\vec k \cdot \vec l \,\, \right)^n \, \frac{C_A}{C_F}
\]
\beq
\times \left\{
C_A\left(\ln\frac{s_0}{\vec k^2}+\frac{2}{\epsilon}-\frac{11}{6}\right)
+\frac{n_f}{3} \right\} \;,
\label{Gvirt-div}
\eeq
and the finite parts,
\[
\left(I_g^V\right)_r=
-\frac{\alpha_s}{2\pi}
\int\limits^1_{\alpha_h}\frac{dx}{x}f_g(x) D_g^h\left(\frac{\alpha_h}{x}\right)
\left(\vec k^{\,\,2} \right)^{\gamma-{n \over 2}}
\left(\vec k \cdot \vec l \,\, \right)^n  \, \frac{C_A}{C_F}
\]
\beq
\times \left\{
C_A\left(\frac{67}{18}-\frac{\pi^2}{2}\right)-\frac{5}{9}n_f
\right\} \;.
\label{Gvirt-fin}
\eeq

For the corrections due to real emissions, one has to consider quark-antiquark
and two-gluon intermediate states,
\beq
I_g^R=I^R_{g,q}+I^R_{g,g} \ .
\eeq

\begin{figure}[tb]
\centering
\includegraphics[scale=0.8]{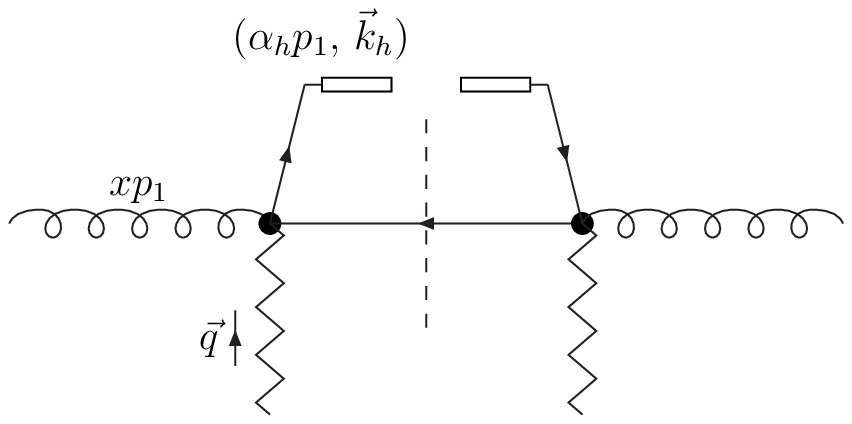}
\caption[]{Diagrammatic representation of the NLO vertex for the identified
hadron production for the case of incoming gluon: real corrections
from quark-antiquark intermediate state, case of quark fragmentation.}
\label{fig:vertex_gluon_NLA_qfrag}
\end{figure}

\subsubsection{Quark-antiquark intermediate state}

The starting point here is the quark-antiquark intermediate state contribution
to the inclusive gluon impact factor ($T_R=1/2$),
\beq
\Phi^{\{Q\bar Q\}}=\Phi_g g^2\vec q^{\,\, 2}\frac{d^{2+2\epsilon} \vec k_1}
{(2\pi)^{3+2\epsilon}}d\beta_1 T_R\left(1-\frac{2\beta_1\beta_2}{1+\epsilon}
\right)\left\{\frac{C_F}{C_A}\frac{1}{\vec k_1^{\, 2} \vec k_2^{\, 2}}
+\beta_1\beta_2\frac{\vec k_1\cdot\vec k_2}{\vec k_1^{\, 2} \vec k_2^{\, 2}
(\vec k_2\beta_1-\vec k_1 \beta_2)^2} \right\}\;,
\eeq
where $\beta_1$ and $\beta_2$ are the relative longitudinal momenta
($\beta_1+\beta_2=1$) and $\vec k_1$ and $\vec k_2$ are the transverse momenta
($\vec k_1+\vec k_2=\vec q\,\,$) of the produced quark and antiquark,
respectively.

We consider quark fragmentation (see Fig.~\ref{fig:vertex_gluon_NLA_qfrag}).
In this case, the ``parent parton'' variables
are  $\vec k=\vec k_1$, $\zeta=\beta_1$ ($\beta_2=\bar \zeta\equiv 1-\zeta$,
$\vec k_2=\vec q-\vec k$). The case when antiquark fragments is essentially
the same. Therefore, the sum over all possible quarks and antiquarks
fragmentation has the form
\[
I^R_{g,q}=
\frac{\alpha_s}{2\pi(4\pi)^\epsilon}\int \frac{d^{2+2\epsilon}\vec q}
{\pi^{1+\epsilon}}\left(\vec q^{\,\,2} \right)^{\gamma-{n \over 2}}
\left(\vec q \cdot \vec l \,\, \right)^n
\int\limits^1_{\alpha_h}
\frac{dx}{x} \int\limits^1_{\frac{\alpha_h}{x}} \frac{d\zeta}{\zeta}
f_g(x) \sum_{a=q,\bar q} D^h_a\left(\frac{\alpha_h}{x \zeta}\right)
\, \frac{C_A}{C_F}
\]
\beq
\times
T_R\left(1-\frac{2\zeta\bar \zeta}{1+\epsilon}\right)\left\{\frac{C_F}{C_A}
\frac{1}{ (\vec q-\vec k)^{ 2}}+\frac{\bar\zeta}{\zeta}
\frac{\vec k\cdot(\vec q-\vec k)}{ (\vec q-\vec k)^{2} (\vec q-\frac{\vec k}
{ \zeta})^2} \right\}\;.
\label{QQquark}
\eeq

We can split this integral into the sum of singular and non-singular parts.
The non-singular contribution of~(\ref{QQquark}) reads
\[
\frac{\alpha_s}{2\pi(4\pi)^\epsilon}
\int\limits^1_{\alpha_h} \frac{dx}{x}
 \int\limits^1_{\frac{\alpha_h}{x}}
\frac{d\zeta}{\zeta} f_g(x) \sum_{a=q,\bar q} D^h_a\left(\frac{\alpha_h}
{x \zeta}\right) \left(\vec k^{\,\,2} \right)^{\gamma+\epsilon-{n \over 2}}
\left(\vec k \cdot \vec l \,\, \right)^n\, \frac{C_A}{C_F}
 T_R\left(1-\frac{2\zeta\bar \zeta}{1+\epsilon}\right)
\frac{\bar \zeta}{\zeta} \]
\beq
\times \int \frac{d^{2+2\epsilon}\vec a}{\pi^{1+\epsilon}}
\left[\left(\vec a^{\, 2} \right)^{\gamma-\frac{n}{2}}
\left(\frac{\vec a \cdot \vec l \,\,}{\vec n \cdot \vec l \,\,} \right)^n
-\zeta^{-2\gamma}\right]\frac{\vec a\cdot \vec n -1}{(\vec a-\vec n)^2
\left(\vec a-\frac{\vec n}{\zeta}\right)^2}\;.
\label{QQquark-reg}
\eeq
Expanding it in $\epsilon$ we obtain
\beq
\left(I^R_{g,q}\right)_r=
\frac{\alpha_s}{2\pi}
\int\limits^1_{\alpha_h} \frac{dx}{x}
 \int\limits^1_{\frac{\alpha_h}{x}}
\frac{d\zeta}{\zeta} f_g(x) \sum_{a=q,\bar q} D^h_a\left(\frac{\alpha_h}
{x \zeta}\right) \left(\vec k^{\,\,2} \right)^{\gamma-{n \over 2}}
\left(\vec k \cdot \vec l \,\, \right)^n\, \frac{C_A}{C_F}\,
\frac{\bar \zeta}{\zeta} \, P_{qg}(\zeta)\, I_3 \; ,
\eeq
where we define the function
\beq
I_3=I_3(n,\gamma,\zeta) =\int \frac{d^{2}\vec a}{\pi}
\frac{\vec a\cdot \vec n -1}{(\vec a-\vec n)^2
\left(\vec a-\frac{\vec n}{\zeta}\right)^2}\left[\left(\vec a^{\, 2}
\right)^{\gamma}e^{in\phi}-\zeta^{-2\gamma}\right]\;.
\eeq

For the singular contribution of~(\ref{QQquark}) we have
\[
\frac{\alpha_s}{2\pi}\frac{\Gamma[1-\epsilon]}{(4\pi)^\epsilon}\frac{1}
{\epsilon}\frac{\Gamma^2(1+\epsilon)}{\Gamma(1+2\epsilon)}
\int\limits^1_{\alpha_h} \frac{dx}{x} f_g(x)
\int\limits^1_{\frac{\alpha_h}{x}} \frac{d\zeta}{\zeta}\sum_{a=q,\bar q}
D^h_a\left(\frac{\alpha_h}{x \zeta}\right) \left(\vec k^{\,\,2}
\right)^{\gamma+\epsilon-{n \over 2}}\left(\vec k \cdot \vec l \,\, \right)^n\,
\frac{C_A}{C_F}
\]
\beq
\times T_R\left(1-\frac{2\zeta\bar \zeta}{1+\epsilon}\right)
\left[
\frac{C_F}{C_A}\frac{\Gamma(1+2\epsilon)\Gamma(\frac{n}{2}-\gamma-\epsilon)
\Gamma(\frac{n}{2}+1+\gamma+\epsilon)}{\Gamma(1+\epsilon)\Gamma(1-\epsilon)
\Gamma(\frac{n}{2}-\gamma)\Gamma(\frac{n}{2}+1+\gamma+2\epsilon)}
+\bar\zeta^{2\epsilon}\zeta^{-2\epsilon-2\gamma} \right]\;.
\label{QQquark-sing}
\eeq
Expanding it in $\epsilon$, we get
\bea
&&
\left(I^R_{g,q}\right)_s=
\frac{\alpha_s}{2\pi}\frac{\Gamma[1-\epsilon]}{\epsilon (4\pi)^\epsilon}
\int\limits^1_{\alpha_h} \frac{dx}{x} \int\limits^1_{\frac{\alpha_h}{x}}
\frac{d\zeta}{\zeta} f_g(x) \sum_{a=q,\bar q}
D^h_a\left(\frac{\alpha_h}{x \zeta}\right)\left(\vec k^{\,\,2}
\right)^{\gamma+\epsilon-{n \over 2}}
\left(\vec k \cdot \vec l \,\, \right)^n\, \frac{C_A}{C_F}
\nonumber \\
&&
\times
\left\{ P_{qg}(\zeta)\left[\frac{C_F}{C_A}+\zeta^{-2\gamma}\right] \right.
\\
&&\left.
+\epsilon \left(2 \zeta \bar\zeta \, T_R \, \left(\frac{C_F}{C_A}
+\zeta^{-2\gamma}\right)
+P_{qg}(\zeta)\, \left(\frac{C_F}{C_A}\, \chi(n,\gamma)+2 \zeta^{-2\gamma}\,
\ln\frac{\bar\zeta}{\zeta}\right)\right)
\right\} \ .
\nonumber
\label{QQquark-div}
\eea

Note that the divergent part of this expression is canceled  by 
the corresponding term of the collinear counterterm~(\ref{c.count.t}).

\begin{figure}[tb]
\centering
\includegraphics[scale=0.8]{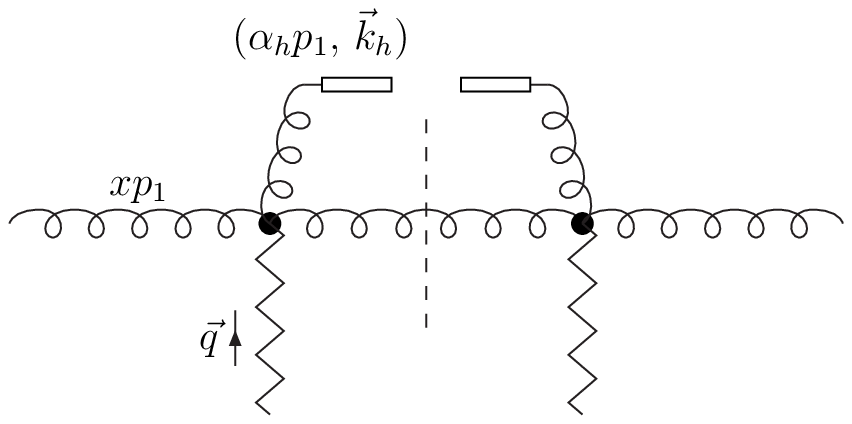}
\caption[]{Diagrammatic representation of the NLO vertex for the identified
hadron production for the case of incoming gluon: real corrections
from the two-gluon intermediate state, case of gluon fragmentation.}
\label{fig:vertex_gluon_NLA_gfrag}
\end{figure}

\subsubsection{Two-gluon intermediate state}

The starting point here is the gluon-gluon intermediate state contribution
to the inclusive gluon impact factor,
\[
\Phi^{\{GG\}}=\Phi_g g^2\vec q^{\,\, 2}\frac{d^{2+2\epsilon} \vec k_1}
{(2\pi)^{3+2\epsilon}}d\beta_1 \frac{C_A}{2}\left[\frac{1}{\beta_1}
+\frac{1}{\beta_2}-2+\beta_1\beta_2\right]
\]
\beq
\times
\left\{\frac{1}{\vec k_1^{\, 2} \vec k_2^{\, 2}}+\frac{\beta_1^2}
{\vec k_1^{\, 2} (\vec k_2\beta_1-\vec k_1 \beta_2)^2}
+\frac{\beta_2^2}{\vec k_2^{\, 2} (\vec k_2\beta_1-\vec k_1 \beta_2)^2}
\right\} \;,
\label{GG}
\eeq
where $\beta_1$ and $\beta_2$ are the relative longitudinal momenta
($\beta_1+\beta_2=1$) and $\vec k_1$ and $\vec k_2$ are the transverse
momenta ($\vec k_1+\vec k_2=\vec q$) of the two produced gluons.

We need to consider the case of a gluon which fragments, the case when the
other gluon fragments being taken into account by a factor 2
(see Fig.~\ref{fig:vertex_gluon_NLA_gfrag}). Thus, we obtain the following
integral:
\[
I_{g,g}^R=\frac{\alpha_s}{2\pi(4\pi)^\epsilon}\int d^{2+2\epsilon}\vec q
\left(\vec q^{\,\,2} \right)^{\gamma-{n \over 2}}
\left(\vec q \cdot \vec l \,\, \right)^n \int\limits^1_{\alpha_h}
\frac{dx}{x} \int\limits^1_{\frac{\alpha_h}{x}} \frac{d\zeta}{\zeta}f_g(x)
D^h_g\left(\frac{\alpha_h}{x \zeta}\right)\, \frac{C_A}{C_F}
\]
\beq
\times C_A\left[\frac{1}{\zeta}+\frac{1}{(1- \zeta)}-2+\zeta\bar\zeta\right]
\left\{\frac{1}{ (\vec q-\vec k)^2}+\frac{1}{ \left(\vec q
-\frac{\vec k}{ \zeta}\right)^2} +\frac{\bar \zeta^2}{\zeta^2}
\frac{ \vec k^2}{(\vec q-\vec k)^2 \left(\vec q -\frac{\vec k}{\zeta}\right)^2}
\right\} \;.
\label{GGgluon}
\eeq

The calculation goes along the same lines as in the Section~\ref{subsub_QG}
(case {\bf b$_2$}). First, we separate the $\zeta\to 1$ singularity,
then we add the BFKL subtraction term. Using~(\ref{zeta0}) one obtains
\[
\left[\frac{1}{\zeta}+\frac{1}{(1- \zeta)}-2+\zeta\bar\zeta\right]
\left\{\frac{1}{ (\vec q-\vec k)^2}+\frac{1}{ \left(\vec q
-\frac{\vec k}{ \zeta}\right)^2} +\frac{\bar \zeta^2}{\zeta^2}
\frac{ \vec k^2}{(\vec q-\vec k)^2 \left(\vec q -\frac{\vec k}{\zeta}\right)^2}
\right\}
\]
\[
=\left[\frac{1}{\zeta}-2+\zeta\bar\zeta\right]
\left\{\frac{1}{ (\vec q-\vec k)^2}+\frac{1}{ \left(\vec q -\frac{\vec k}
{\zeta}\right)^2} +\frac{\bar \zeta^2}{\zeta^2}\frac{ \vec k^2}
{(\vec q-\vec k)^2 \left(\vec q -\frac{\vec k}{ \zeta}\right)^2}\right\}
\]
\[
+\frac{1}{(1- \zeta)}\left\{ \frac{1}{ \left(\vec q -\frac{\vec k}{ \zeta}
\right)^2}-\frac{1}{ (\vec q-\vec k)^2}+\frac{\bar \zeta^2}{\zeta^2}
\frac{ \vec k^2}{(\vec q-\vec k)^2 \left(\vec q -\frac{\vec k}{\zeta}\right)^2}
\right\}+\frac{1}{(1- \zeta)}\frac{2}{ (\vec q-\vec k)^2}
\]
\[
\to\left[\frac{1}{\zeta}-2+\zeta\bar\zeta\right]
\left\{\frac{1}{ (\vec q-\vec k)^2}+\frac{1}{ \left(\vec q -\frac{\vec k}
{\zeta}\right)^2} +\frac{\bar \zeta^2}{\zeta^2}\frac{ \vec k^2}
{(\vec q-\vec k)^2 \left(\vec q -\frac{\vec k}{ \zeta}\right)^2}\right\}
\]
\[
+\frac{1}{(1-\zeta)}\left\{\frac{1}{\left(\vec q -\frac{\vec k}{\zeta}
\right)^2}-\frac{1}{ (\vec q-\vec k)^2}+\frac{\bar \zeta^2}{\zeta^2}
\frac{ \vec k^2}{(\vec q-\vec k)^2 \left(\vec q -\frac{\vec k}{\zeta}\right)^2}
\right\}+\frac{1}{(1- \zeta)_+}\frac{2}{ (\vec q-\vec k)^2}
\]
\[
+\delta(1-\zeta)\frac{1}{ (\vec q-\vec k)^2}\ln\frac{s_0}{(\vec q-\vec k)^2 }
\;.
\]

Again, we can split the result into the sum of  singular and non-singular
parts. The non-singular contribution differs from~(\ref{QGquark-CF-reg}) only
by a factor, it  reads
\[
\frac{\alpha_s}{2\pi(4\pi)^\epsilon}
 \int\limits^1_{\alpha_h} \frac{dx}{x} \int\limits^1_{\frac{\alpha_h}{x}}
\frac{d\zeta}{\zeta} f_g(x) D^h_g\left(\frac{\alpha_h}{x \zeta}\right)
\left(\vec k^{\,\,2} \right)^{\gamma+\epsilon-{n \over 2}}
\left(\vec k \cdot \vec l \, \right)^n
\frac{\bar \zeta^2}{\zeta^2}\left[\frac{1}{\zeta}+\frac{1}{(1- \zeta)}-2
+\zeta\bar\zeta\right]
 \frac{C_A}{C_F}
\]
\beq
\times
C_A\int \frac{d^{2+2\epsilon}\vec a}{\pi^{1+\epsilon}}
\frac{1}{(\vec a-\vec n)^2+\left(\vec a-\frac{\vec n}{\zeta}\right)^2}
 \left[\frac{\left(\vec a^{\,\,2} \right)^{\gamma-{n \over 2}}
\left(\frac{\vec a \cdot \vec l }{\vec n \cdot \vec l \,\, }\right)^n-1}
{(\vec a-\vec n)^2}+
\frac{\left(\vec a^{\,\,2} \right)^{\gamma-{n \over 2}}
\left(\frac{\vec a \cdot \vec l }{\vec n \cdot \vec l \,\, }\right)^n
-\zeta^{-2\gamma}}{(\vec a-\frac{\vec n}{\zeta})^2}
\right]\; .
\label{GG-reg}
\eeq

Expanding it in $\epsilon$ we obtain
\[
\left(I^R_{g,g}\right)_r=
\frac{\alpha_s}{2\pi}
 \int\limits^1_{\alpha_h} \frac{dx}{x} \int\limits^1_{\frac{\alpha_h}{x}}
\frac{d\zeta}{\zeta} f_g(x) D^h_g\left(\frac{\alpha_h}{x \zeta}\right)
\left(\vec k^{\,\,2} \right)^{\gamma-{n \over 2}}
\left(\vec k \cdot \vec l \, \right)^n
\]
\beq
\times
\frac{\bar \zeta^2}{\zeta^2}\left[\frac{1}{\zeta}+\frac{1}{(1- \zeta)}-2
+\zeta\bar\zeta\right]
 \frac{C_A}{C_F} C_A I_2\; .
\label{GG-reg1}
\eeq

For the singular contribution we obtain
\[
\frac{\alpha_s}{2\pi}\frac{\Gamma[1-\epsilon]}{(4\pi)^\epsilon}
\frac{1}{\epsilon}\frac{\Gamma^2(1+\epsilon)}{\Gamma(1+2\epsilon)}
\int\limits^1_{\alpha_h} \frac{dx}{x}  \int\limits^1_{\frac{\alpha_h}{x}}
\frac{d\zeta}{\zeta}
f_g(x) D^h_g\left(\frac{\alpha_h}{x \zeta}\right)
\left(\vec k^{\,\,2} \right)^{\gamma+\epsilon-{n \over 2}}
\left(\vec k \cdot \vec l \,\, \right)^n \, C_A \, \frac{C_A}{C_F}
\]
\[
\times\!\left\{
\left[\frac{1}{\zeta}+\frac{1}{(1-\zeta)}-2+\zeta\bar\zeta\right]
\left(\frac{\bar \zeta}{\zeta}\right)^{2\epsilon}\!
\left(1+\zeta^{-2\gamma}\right)+
\frac{\Gamma(1+2\epsilon)\Gamma(\frac{n}{2}-\gamma-\epsilon)
\Gamma(\frac{n}{2}+1+\gamma+\epsilon)}
{\Gamma(1+\epsilon)\Gamma(1-\epsilon)\Gamma(\frac{n}{2}-\gamma)
\Gamma(\frac{n}{2}+1+\gamma+2\epsilon)}
\right.
\]
\[
\times \left[
\delta(1-\zeta)\left(\ln\frac{s_0}{\vec k^2}
+\psi\left(\frac{n}{2}-\gamma-\epsilon\right)
+\psi\left(1+\gamma+\frac{n}{2}+2\epsilon\right)-\psi(\epsilon)-\psi(1)\right)
+\frac{2}{(1-\zeta)_+}\right.
\]
\beq
\left.\left.+\frac{
\left(\zeta^{-2\epsilon-2\gamma}-1\right)}{(1-\zeta)}
+\left[\frac{1}{\zeta}-2+\zeta\bar\zeta\right]
\left(1+\zeta^{-2\epsilon-2\gamma}\right)\right]\right\} \;.
\label{GG-sing}
\eeq
Expanding this result in $\epsilon$, we get for the divergent contribution
\[
\frac{\alpha_s}{2\pi}\frac{\Gamma[1-\epsilon]}{\epsilon (4\pi)^\epsilon}
\frac{\Gamma^2(1+\epsilon)}{\Gamma(1+2\epsilon)}
\int\limits^1_{\alpha_h} \frac{dx}{x}  \int\limits^1_{\frac{\alpha_h}{x}}
\frac{d\zeta}{\zeta} f_g(x) D^h_g\left(\frac{\alpha_h}
{x \zeta}\right)\left(\vec k^{\,\,2} \right)^{\gamma+\epsilon-{n \over 2}}
\left(\vec k \cdot \vec l \,\, \right)^n\, \frac{C_A}{C_F}
\]
\[
\times C_A \left\{
2 \left[\frac{1}{\zeta}+\frac{1}{(1- \zeta)_+}-2+\zeta\bar\zeta\right]
\left(1+\zeta^{-2\gamma}\right)+
\delta(1-\zeta)\left(\ln\frac{s_0}{\vec k^2}+\frac{2}{\epsilon}\right)
\right\}
\]
\[
=\frac{\alpha_s}{2\pi}\frac{\Gamma[1-\epsilon]}{\epsilon (4\pi)^\epsilon}
\frac{\Gamma^2(1+\epsilon)}{\Gamma(1+2\epsilon)}
\int\limits^1_{\alpha_h} \frac{dx}{x}  \int\limits^1_{\frac{\alpha_h}{x}}
\frac{d\zeta}{\zeta}
f_g(x) D^h_g\left(\frac{\alpha_h}{x \zeta}\right)\left(\vec k^{\,\,2}
\right)^{\gamma+\epsilon-{n \over 2}}
\left(\vec k \cdot \vec l \,\, \right)^n\, \frac{C_A}{C_F}
\]
\beq
\times \left\{P_{gg}(\zeta)\left(1+\zeta^{-2\gamma}\right)+
\delta(1-\zeta)\left[C_A\left(\ln\frac{s_0}{\vec k^2}
+\frac{2}{\epsilon}-\frac{11}{3}\right)+\frac{2n_f}{3}\right]\right\} \;.
\label{GG-div}
\eeq

Finally, the $\epsilon$ expansion of the divergent part has the form
\[
\left(I^R_{g,g}\right)_s=
\frac{\alpha_s}{2\pi}\frac{\Gamma[1-\epsilon]}{\epsilon(4\pi)^\epsilon}
\frac{\Gamma^2(1+\epsilon)}{\Gamma(1+2\epsilon)}
\int\limits^1_{\alpha_h} \frac{dx}{x}  \int\limits^1_{\frac{\alpha_h}{x}}
\frac{d\zeta}{\zeta} f_g(x) D^h_g\left(\frac{\alpha_h}
{x \zeta}\right)\left(\vec k^{\,\,2} \right)^{\gamma+\epsilon-{n \over 2}}
\left(\vec k \cdot \vec l \,\, \right)^n\, \frac{C_A}{C_F}
\]
\[
\times \left\{P_{gg}(\zeta)\left(1+\zeta^{-2\gamma}\right)+
\delta(1-\zeta)\left[C_A\left(\ln\frac{s_0}{\vec k^2}
+\frac{2}{\epsilon}-\frac{11}{3}\right)+\frac{2n_f}{3}\right]\right.
\]
\[
+\, \epsilon \, C_A\left[\delta(1-\zeta)\left(\chi(n,\gamma)\ln\frac{s_0}
{\vec k^2}
+\frac{1}{2}\left(\psi^\prime\left(1+\gamma+\frac{n}{2}\right)
-\psi^\prime\left(\frac{n}{2}-\gamma\right)
-\chi^2(n,\gamma)\right)\right)\right.
\]
\[
+\left(\frac{1}{\zeta}+\frac{1}{(1-\zeta)_+}-2+\zeta\bar\zeta\right)
\left(\chi(n,\gamma)(1+\zeta^{-2\gamma})-2(1+2\zeta^{-2\gamma})\ln\zeta\right)
\]
\beq
\left.\left.
+2(1+\zeta^{-2\gamma})
\left(\left(\frac{1}{\zeta}-2+\zeta\bar\zeta\right) \ln\bar\zeta
+\left(\frac{\ln(1-\zeta)}{1-z}\right)_+\right)\right]\right\}\;.
\label{GG-fin}
\eeq
The term in the divergent contribution~(\ref{GG-fin}) proportional to $P_{gg}$
cancels with the corresponding term of the collinear
counterterm~(\ref{c.count.t}), while the term in~(\ref{GG-fin}) proportional
to $\delta(1-\zeta)$ cancels in the sum with the singular part of the virtual
corrections~(\ref{Gvirt-div}). The uncanceled divergence in the virtual
corrections~(\ref{Gvirt-div}) vanishes after the QCD charge
renormalization~(\ref{charge-ren}).

Summarizing, all the infrared and ultraviolet divergences arisen in the
calculation have disappeared after taking into account PDFs and FFs
renormalization and QCD charge renormalization. Collecting all intermediate
contributions we obtain the final result for the identified hadron NLO impact
factor, which reads
\beq
\vec k^{\, 2}_h \,\,\frac{d\Phi^h(\nu,n)}
{d\alpha_h d^2\vec k_h}=2\,\alpha_s(\mu_R)\sqrt{\frac{C_F}{C_A}}\,
\left(\vec k_h^{\,\,2}
\right)^{\gamma-{n \over 2}}\left(\vec k_h \cdot \vec l \,\, \right)^n
\label{final}
\eeq
\[
\times
\left\{ \ \
\int\limits^1_{\alpha_h}\frac{dx}{x} \left(\frac{x}{\alpha_h}\right)^{2\gamma}
\left[ \frac{C_A}{C_F}f_g(x)
D_g^h\left(\frac{\alpha_h}{x}\right) +\sum_{a=q,\bar q}f_a(x)
D_a^h\left(\frac{\alpha_h}{x}\right)\right]
\right.
\]
\[
+\frac{\alpha_s\left(\mu_R\right)}{2\pi}\!\int\limits^1_{\alpha_h}\frac{dx}{x}
\!\int\limits^1_{\frac{\alpha_h}{x}}
\frac{d\zeta}{\zeta}\left(\frac{x\, \zeta}{\alpha_h}\right)^{2\gamma}
\!\!\left[\frac{C_A}{C_F}f_g(x)
D_g^h\left(\frac{\alpha_h}{x\zeta}\right)C_{gg}\left(x,\zeta\right)
+\sum_{a=q,\bar q}f_a(x)
D_a^h\left(\frac{\alpha_h}{x\zeta}\right)C_{qq}\left(x,\zeta\right)
\right.
\]
\[
\left.\left.
+\sum_{a=q,\bar q}f_a(x)
D_{g}^h\left(\frac{\alpha_h}{x\zeta}\right)C_{qg}\left(x,\zeta\right)
+\frac{C_A}{C_F} f_{g}(x)\sum_{a=q,\bar q}
D_a^h\left(\frac{\alpha_h}{x\zeta}\right)C_{gq}\left(x,\zeta\right) \right]
\right\}\;.
\]

The results for the NLO coefficient functions read
\bea
&&
C_{gg}\left(x,\zeta\right) =  P_{gg}(\zeta)\left(1+\zeta^{-2\gamma}\right)
\ln \left( \frac {\vec k_h^2 x^2 \zeta^2 }{\mu_F^2 \alpha_h^2}\right)
-\frac{\beta_0}{2}\ln \left( \frac {\vec k_h^2 x^2 \zeta^2 }
{\mu^2_R \alpha_h^2}\right)
\\
&&
+ \, \delta(1-\zeta)\left[C_A \ln\left(\frac{s_0 \, \alpha_h^2}{\vec k^2_h \,
x^2 }\right) \chi(n,\gamma)
- C_A\left(\frac{67}{18}-\frac{\pi^2}{2}\right)+\frac{5}{9}n_f
\right.
\nonumber \\
&&
\left.
+\frac{C_A}{2}\left(\psi^\prime\left(1+\gamma+\frac{n}{2}\right)
-\psi^\prime\left(\frac{n}{2}-\gamma\right)
-\chi^2(n,\gamma)\right) \right]
\nonumber \\
&&
+ \, C_A \left(\frac{1}{\zeta}+\frac{1}{(1-\zeta)_+}-2+\zeta\bar\zeta\right)
\left(\chi(n,\gamma)(1+\zeta^{-2\gamma})-2(1+2\zeta^{-2\gamma})\ln\zeta
+\frac{\bar \zeta^2}{\zeta^2}I_2\right)
\nonumber \\
&&
+ \, 2 \, C_A (1+\zeta^{-2\gamma})
\left(\left(\frac{1}{\zeta}-2+\zeta\bar\zeta\right) \ln\bar\zeta
+\left(\frac{\ln(1-\zeta)}{1-\zeta}\right)_+\right) \ ,
\nonumber
\eea

\bea
&&
C_{gq}\left(x,\zeta\right) =  P_{qg}(\zeta)\left(\frac{C_F}{C_A}
+\zeta^{-2\gamma}\right)
\ln \left( \frac {\vec k_h^2 x^2 \zeta^2 }{\mu_F^2 \alpha_h^2}\right)
+ \, 2 \, \zeta \bar\zeta \, T_R \, \left(\frac{C_F}{C_A}+\zeta^{-2\gamma}
\right)
\\
&&
+\, P_{qg}(\zeta)\, \left(\frac{C_F}{C_A}\, \chi(n,\gamma)+2 \zeta^{-2\gamma}\,
\ln\frac{\bar\zeta}{\zeta} + \frac{\bar \zeta}{\zeta}I_3\right) \ ,
\nonumber
\eea

\bea
&&
C_{qg}\left(x,\zeta\right) =  P_{gq}(\zeta)\left(\frac{C_A}{C_F}
+\zeta^{-2\gamma}\right)
\ln \left( \frac {\vec k_h^2 x^2 \zeta^2 }{\mu_F^2 \alpha_h^2}\right)
+ \zeta\left(C_F\zeta^{-2\gamma}+C_A\right)
\\
&&
+ \, \frac{1+\bar \zeta^2}{\zeta}\left[C_F\zeta^{-2\gamma}(\chi(n,\gamma)
-2\ln\zeta)
+2C_A\ln\frac{\bar \zeta}{\zeta} + \frac{\bar \zeta}{\zeta}I_1\right]
 \ ,
\nonumber
\eea

\bea
&&
C_{qq}\left(x,\zeta\right) =  P_{qq}(\zeta)\left(1+\zeta^{-2\gamma}\right)
\ln \left( \frac {\vec k_h^2 x^2 \zeta^2 }{\mu_F^2 \alpha_h^2}\right)
-\frac{\beta_0}{2}\ln \left( \frac {\vec k_h^2 x^2 \zeta^2 }{\mu^2_R
\alpha_h^2}\right)
\label{final-e}
\\
&&
+ \, \delta(1-\zeta)\left[C_A \ln\left(\frac{s_0 \, \alpha_h^2}{\vec k^2_h \,
x^2 }\right) \chi(n,\gamma)
+ C_A\left(\frac{85}{18}+\frac{\pi^2}{2}\right)-\frac{5}{9}n_f - 8\, C_F
\right.
\nonumber \\
&&
\left.
+\frac{C_A}{2}\left(\psi^\prime\left(1+\gamma+\frac{n}{2}\right)
-\psi^\prime\left(\frac{n}{2}-\gamma\right)
-\chi^2(n,\gamma)\right) \right] + \, C_F \,\bar \zeta\,
(1+\zeta^{-2\gamma})
\nonumber \\
&&
+  \left(1+\zeta^2\right)\left[C_A (1+\zeta^{-2\gamma})\frac{\chi(n,\gamma)}
{2(1-\zeta )_+}
+\left(C_A-2\, C_F(1+\zeta^{-2\gamma})\right)\frac{\ln \zeta}{1-\zeta}
\right]
\nonumber\\
&&
+\, \left(C_F-\frac{C_A}{2}\right)\left(1+\zeta^2\right)
\left[2(1+\zeta^{-2\gamma})\left(\frac{\ln (1-\zeta)}{1-\zeta}\right)_+
+ \frac{\bar \zeta}{\zeta^2}I_2\right]\ .
\nonumber
\eea

For the $I_{1,2,3}$ functions we obtain the following results:
\beq
I_2=\frac{\zeta^2}{\bar \zeta^2}\left[
\zeta\left(\frac{{}_2F_1(1,1+\gamma-\frac{n}{2},2+\gamma-\frac{n}{2},\zeta)}
{\frac{n}{2}-\gamma-1}-
\frac{{}_2F_1(1,1+\gamma+\frac{n}{2},2+\gamma+\frac{n}{2},\zeta)}{\frac{n}{2}+
\gamma+1}\right)\right.
\eeq
$$
\left.
+\zeta^{-2\gamma}
\left(\frac{{}_2F_1(1,-\gamma-\frac{n}{2},1-\gamma-\frac{n}{2},\zeta)}
{\frac{n}{2}+\gamma}-
\frac{{}_2F_1(1,-\gamma+\frac{n}{2},1-\gamma+\frac{n}{2},\zeta)}{\frac{n}{2}
-\gamma}\right)
\right.
$$
$$
\left.
+\left(1+\zeta^{-2\gamma}\right)\left(\chi(n,\gamma)-2\ln \bar \zeta \right)
+2\ln{\zeta}\right]\;,
$$

\beq
I_1=\frac{\bar \zeta}{2\zeta}I_2+\frac{\zeta}{\bar \zeta}\left[
\ln \zeta+\frac{1-\zeta^{-2\gamma}}{2}\left(\chi(n,\gamma)-2\ln \bar \zeta
\right)\right]\;,
\eeq

\beq
I_3=\frac{\bar \zeta}{2\zeta}I_2-\frac{\zeta}{\bar \zeta}\left[
\ln \zeta+\frac{1-\zeta^{-2\gamma}}{2}\left(\chi(n,\gamma)-2\ln \bar \zeta
\right)\right]\;.
\eeq

Using the following property of the hypergeometric function,
$$
{}_2F_1(1,a,a+1,\zeta)=a(\psi(1)-\psi(a)-\ln\bar \zeta)
+{\cal O}(\bar \zeta\ln \bar \zeta)\;,
$$
one can easily see that
$$
I_2={\cal O}\left( \ln \bar \zeta\right)\, ,
\quad I_1={\cal O}(\ln \bar \zeta ) \, ,\quad I_3={\cal O}(\ln \bar \zeta )\,,
$$
which implies that the integral over $\zeta$ in~(\ref{final}) is convergent
in the upper limit.

\section{Summary}

In this paper we have calculated the NLO vertex (impact factor) for the forward
production of an identified hadron from an incoming quark or gluon, emitted
by a proton. This is a necessary ingredient for the calculation of the
hard inclusive production of a pair of rapidity-separated identified hadrons
in proton collisions~(\ref{process}). This process, similarly to the
production of Mueller-Navelet jets, can be studied at the LHC hadron collider.

Another natural application of the obtained identified hadron production vertex
could be the NLA BFKL description of inclusive forward hadron production
process in DIS,
\beq
e(p_1)+p(p_2)\to h(k)+X \ ,
\label{process1}
\eeq
where in the low-$x$ event the hadron $h(k)$ with high transverse momentum is
detected in the fragmentation region of incoming proton $p(p_2)$. Data
for such reaction in the case of forward $\pi^0$-production were published by
the H1 collaboration at HERA~\cite{Aktas:2004rb}.

At the basis of our calculation of the hard part of the vertex was the
definition of NLO BFKL parton impact factors; then the collinear factorization
with the PDFs of the incoming partons and with the FF for the production
of the identified hadron (in the $\overline{\rm{MS}}$ scheme) was suitably
considered.

We have presented our result for the vertex in the so called
$(\nu,n)$-representation, which is the most convenient one in view of the
numerical determination of the cross section for the production of
a pair of rapidity-separated identified hadrons along the same lines as
in Ref.~\cite{mesons}.

We have explicitly verified that soft and virtual infrared divergences cancel
each other, whereas the infrared collinear ones are compensated by the PDFs'
and FFs' renormalization counterterms, the remaining ultraviolet divergences
being taken care of by the renormalization of the QCD coupling.

In our approach the energy scale $s_0$ is an arbitrary parameter, that need not
be fixed at any definite scale. The dependence on $s_0$ will disappear
in the next-to-leading logarithmic approximation in any physical cross
section in which the identified hadron production  vertices are used. Indeed,
our result for the NLO vertex, given by Eqs.~(\ref{final})-(\ref{final-e}),
contains contributions $\sim \ln(s_0)$ and these terms
are proportional to the LO quark and gluon vertices multiplied by the BFKL
kernel eigenvalue $\chi(n,\nu)$. This fact guarantees the independence
of the identified hadrons~(\ref{process}) or  single 
hadron~({\ref{process1}}) cross section on $s_0$ within the next-to-leading
logarithmic approximation.
However, the dependence on this energy scale will survive in terms beyond
this approximation and will provide a parameter to be optimized with the
method adopted in Refs.~\cite{mesons}.

\section*{Acknowledgements}

D.I. thanks the Dipartimento di Fisica dell'U\-ni\-ver\-si\-t\`a della Calabria
and the Istituto Nazio\-na\-le di Fisica Nucleare (INFN), Gruppo collegato di
Cosenza, for the warm hospitality and the financial support. This work was
also supported in part by the grants and RFBR-11-02-00242 and NSh-3810.2010.2.

\appendix
\section{Useful integrals}

We give here some useful integrals:
\beq
{\rm \bullet}\;\;
\int\frac{d^{2+2\epsilon}\vec k^\prime}{\vec k^{\prime 2}}
\left(\frac{1}{\vec k^{\prime 2}+(\vec k^\prime-\vec k \,)^2}\right) =
\frac{1}{2}\int\frac{d^{2+2\epsilon}\vec k^\prime}{\vec k^{\prime 2}
(\vec k^\prime-\vec k)^2}=\pi^{1+\epsilon}
\left(\vec k^{\, 2}\right)^{\epsilon -1}
\frac{\Gamma(1-\epsilon) \Gamma^2(1+\epsilon)}
{\epsilon\, \Gamma(1+2\epsilon)}\;,
\label{I0}
\eeq

\beq
{\rm \bullet}\;\;
\int \frac{d^{2+2\epsilon}\vec k^\prime (\vec k^{\prime 2})^\alpha}
{(\vec k-\vec k^\prime)^2} =
\pi^{1+\epsilon}\left(\vec k^2\right)^{\alpha + \epsilon}
\frac{\Gamma(-\epsilon-\alpha)}{\Gamma(-\alpha)}\frac{\Gamma(\epsilon)\,
\Gamma(1+\epsilon+\alpha)}{\Gamma(1+\alpha+2\epsilon)}\;.
\label{I1}
\eeq

\beq
{\rm \bullet}\;\;
\int\limits^{2\pi}_0 d\phi\frac{\cos n\phi}{a^2-2 a b \cos\phi +b^2 }
=\frac{2\pi}{b^2-a^2}\left(\frac{a}{b}\right)^n\ , \quad a<b
\label{I4}
\eeq

In the integrals below, $\vec l^2=0$ is assumed:
\beq
{\rm \bullet}\;\;
\int \frac{\displaystyle d^{2+2\epsilon} \vec k^\prime
(\vec k^{\prime 2})^\alpha (\vec k^\prime \cdot \vec l \, )^\beta}
{\displaystyle(\vec k-\vec k^\prime)^2} =
\pi^{1+\epsilon}  \left( \vec k\cdot \vec l\, \right)^{\beta}
\left(\vec k^2\right)^{\alpha +\epsilon}
\frac{\displaystyle \Gamma(-\alpha-\epsilon)}{\displaystyle
\Gamma(-\alpha)}\frac{\displaystyle \Gamma(\epsilon)\,
\Gamma(1+\epsilon+\alpha+\beta)}{\displaystyle
\Gamma(1+\alpha+\beta+2\epsilon)}\;,
\label{I2}
\eeq

\beq
{\rm \bullet}\;\;
{\displaystyle \int} \frac{\displaystyle d^{2+2\epsilon} \vec k^\prime
\ln (\vec k-\vec k^\prime)^2  (\vec k^{\prime 2})^\alpha
(\vec k^\prime \cdot \vec l\, )^\beta}{\displaystyle (\vec k-\vec k^\prime)^2}
= \pi^{1+\epsilon}  \left( \vec k\cdot \vec l\, \right)^{\beta}
\left(\vec k^2\right)^{\alpha +\epsilon}
\eeq
\[
\times \frac{\displaystyle \Gamma(-\alpha-\epsilon)}{\displaystyle
\Gamma(-\alpha)}\frac{\displaystyle \Gamma(\epsilon)\,
\Gamma(1+\epsilon+\alpha+\beta)}
{\displaystyle \Gamma(1+\alpha+\beta+2\epsilon)}
\]
\[
\times\left\{
\ln \vec k^2 +\psi(\epsilon)+ \psi(1)- \psi(-\alpha-\epsilon)
-\psi(1+\alpha+\beta+2\epsilon) \right\} \,.
\label{I3}
\]

\end{document}